\documentclass[a4paper,12pt]{article}

\usepackage{amsmath, amssymb, amsthm, bm, bigints}

\usepackage[T1]{fontenc}
\usepackage[utf8]{inputenc}

\usepackage{graphicx}
\usepackage{float}
\usepackage{caption}
\usepackage{subcaption}
\usepackage{rotating}
\usepackage{xcolor}

\usepackage[margin=2cm]{geometry}  % Only define once

\usepackage[affil-it]{authblk}

\usepackage{setspace}
\usepackage{booktabs}
\usepackage{multirow}
\usepackage{array}
\usepackage{tabularx}
\usepackage{enumitem}

\usepackage{hyperref}

\usepackage[ruled,vlined]{algorithm2e}

\usepackage{placeins}
\usepackage{appendix}
\usepackage{lineno}
\usepackage{changepage}
\usepackage{soul}
\usepackage{xfrac}

\usepackage{geometry}
\usepackage{longtable}
\usepackage[table]{xcolor}
\usepackage{array}

\usepackage{nomencl}

\title{pyMEAL: A Multi-Encoder Augmentation-Aware-Learning Toolbox for Robust Medical Image Translation}

\author[1]{Abdul-mojeed Olabisi Ilyas \thanks{Corresponding author: \href{mailto:amoilyas@hkcoche.org}{amoilyas@hkcoche.org}}}
\author[1]{Adeleke Maradesa \thanks{Corresponding author: \href{mailto:amaradesa@hkcoche.org}{amaradesa@hkcoche.org}}}

\author[1,5]{Jamal Banzi}
\author[6,7,8]{Jianpan  Huang}
\author[6,7,8]{Henry K.F. Mak}
\author[1,2,3,4]{Kannie W.Y. Chan\thanks{Corresponding author: \href{mailto:kannie@hkcoche.org}{kannie@hkcoche.org}}}

\affil[1]{Hong Kong Centre for Cerebro-Cardiovascular Health Engineering (COCHE), Hong Kong, China}
\affil[2]{Department of Biomedical Engineering, City University of Hong Kong, Hong Kong, China}
\affil[3]{Russell H. Morgan Department of Radiology and Radiological Science, The Johns Hopkins University School of Medicine, Baltimore, MD, USA}
\affil[4]{City University of Hong Kong Shenzhen Research Institute, Shenzhen, China}

\affil[5]{Department of Informatics, Sokoine University of Agriculture, Chuo Kikuu Morogoro, Tanzania}
\affil[6]{Department of Diagnostic Radiology, The University of Hong Kong, Hong Kong, China}
\affil[7]{State Key Laboratory of Brain and Cognitive Sciences, The University of Hong Kong, Hong Kong}
\affil[8]{Alzheimer's Disease Research Network, The University of Hong Kong, Hong Kong}

\date{}
\begin{document}

\maketitle

\begin{abstract}

Medical imaging is critical for clinical diagnosis, yet the adoption of advanced AI-driven imaging methods remains challenged by patient variability, image artifacts, and limited robustness across acquisition conditions. Although deep learning has transformed medical image analysis, 3D imaging tasks continue to suffer from data scarcity and variability arising from scanner differences, acquisition protocols, and patient motion. Conventional data augmentation typically relies on a single transformation pipeline, overlooking augmentation-specific characteristics and limiting effective representation learning.

To address these challenges, we propose a Multi-Encoder Augmentation-Aware Learning (MEAL) framework, which processes multiple augmentation variants through dedicated encoder pathways. Three fusion strategies, encoder concatenation (CC), fusion layer (FL), and an adaptive controller block (BD), are introduced to integrate augmentation-specific features prior to decoding. Among these, MEAL-BD preserves augmentation-aware representations through dynamic feature weighting, enabling improved robustness to clinically relevant variability.

We evaluate MEAL in a CT-to-T1-weighted MRI translation case study, where such translation is particularly relevant in settings in which MRI is unavailable, contraindicated, or delayed, including emergency imaging and resource-limited clinical environments. Across unseen and predefined test data, MEAL-BD consistently outperformed competing approaches under both geometric perturbations and non-augmented conditions, achieving higher peak signal-to-noise ratio (PSNR) and structural similarity index measure (SSIM). By prioritizing structural fidelity over perceptual realism, MEAL is designed to support clinical interpretation and downstream analysis rather than directly replacing diagnostic MRI, advancing robust and clinically meaningful medical image translation.

\end{abstract}


\clearpage
\section{Introduction}

Medical imaging has revolutionized modern diagnostics, enabling precise disease detection~\cite{hussain2022modern} and treatment planning~\cite{fountzilas2025convergence, wagner1991advances}. Deep learning (DL) has played a transformative role in this domain, excelling in tasks such as lesion detection~\cite{attariwala2013whole}, segmentation~\cite{cossio2023augmenting,ostertagova2014methodology}, and anomaly classification~\cite{alshardan2024leveraging}. These advances are rooted in DL’s ability to learn hierarchical feature representations that align well with complex patterns in biomedical data~\cite{chaari2025hybrid, alzubaidi2021review}. While DL has further enhanced these capabilities through automated image analysis, its application to 3D medical imaging remains affected by critical challenges such as limited availability of high-quality data, variability in acquisition protocols, scanner heterogeneity, and patient motion. Consequently, these factors compromise the generalizability of the model ~\cite{elton2019deep, huang2023self}, particularly in clinical settings where robustness to real-world variability is crucial. 

Data augmentation, a foundational technique in DL, expands dataset through transformations such as rotation, flipping, scaling, and intensity shifts~\cite{mumuni2022data, alomar2023data}. However, in medical imaging, where anatomical precision is critical, it must be applied with more caution than in the natural image domains~\cite{yang2023survey, garcea2023data}. Traditional augmentation methods, such as geometric transformations or noise perturbation, partly mitigate data scarcity but rely on uniform, single-pipeline transformations~\cite{bosquet2023full}. However, it overlooks the unique value of individual augmentations and may introduce unrealistic artifacts when overused (e.g., excessive noise that distorts tumor boundaries in CT scans~\cite{kaji2019overview}), ultimately affecting model performance~\cite{shorten2019survey}. Recent studies show that random or overly aggressive augmentations can obscure critical diagnostic features, particularly in modalities like MRI and CT where fine-tissue contrast is essential~\cite{kebaili2023deep, chlap2021review}. In some cases, such distortions may not only degrade performance, but also introduce clinically misleading patterns~\cite{abdollahi2020data, cossio2023augmenting}.

The need for augmentation strategies that generalize while preserving medically significant patterns has spurred interest in augmentation-aware learning~\cite{goceri2023medical,przewikezlikowski2024augmentation,kim2025augward}. Frameworks like AugSelf~\cite{przewikezlikowski2024augmentation} and CASSLE~\cite{trzcinski2024zero} use augmentation-specific encoding or self-supervision to retain clinically relevant transformations, but often face challenges such as increased architectural complexity and task-specific overfitting. Recent advances, such as Generative Adversarial Networks ~\cite{sandfort2019data} and hybrid approaches like PixMed-Enhancer~\cite{rasool2025pixmed}, are successful for medical image translation. However, they struggle to leverage the augmentation-induced diversity for anatomical feature learning, particularly in cross-modal translation tasks requiring protocol invariance. Current methods often regard augmentations as noisy variants rather than complementary views, limiting the analysis of transformation-specific features vital for clinical generalizability~\cite{kim2025augward,cossio2023augmenting,przewikezlikowski2024augmentation}. 

Therefore, we propose a new Multi-encoder Augmentation-Aware Learning (MEAL\footnote{The framework may also be referred to as pyMEAL, highlighting its Python-based implementation for medical image translation.}) framework that interprets data augmentation as the generation of diverse anatomical views. MEAL employs parallel encoder pathways for each defined augmentation type, capturing transformation-specific features while maintaining augmentation provenance. A fusion controller with attention-guided softmax weighting dynamically integrates these features into a unified representation for decoding.
This transformative approach addresses key limitations in medical image-to-image translation by transforming augmentation from simple dataset inflation into an anatomical disentanglement framework. It mitigates performance degradation under extreme augmentation through discriminative feature fusion and eliminates scanner bias through invariant representation learning. By treating augmentations as learnable physiological features, MEAL enables augmentation-aware learning that generalizes effectively across diverse scanner settings~\cite{yang2020mri,sandfort2019data}.

We evaluated our framework on CT-to-T1-weighted MRI translation, a critical task in settings where MRI is unavailable but high soft-tissue contrast is essential~\cite{choo2024slice}, such as brain tumor delineation~\cite{chen2019differentiation} in resource-limited environments. While CT is widely accessible, its limited soft-tissue contrast hinders diagnostic effectiveness~\cite{chen2019differentiation}. Existing translation methods often struggle to generalize under real-world conditions, including motion artifacts and protocol variability~\cite{bearcroft2007imaging}. Our framework addresses these challenges by jointly optimizing augmentation-specific encoders with a shared decoder, enabling robust synthesis of MRI-like images from heterogeneous CT inputs. Lastly, our experiments show that MEAL outperforms state-of-the-art single-stream and multi-modal baselines in both SSIM and PSNR. By treating augmentations as complementary views rather than perturbations, MEAL sets a new benchmark for robustness in medical image analysis. Its adaptability extends beyond synthesis to tasks such as segmentation, registration, and real-time clinical decision support, bridging the gap between controlled research settings and clinical variability. Thus, MEAL emerges as a generalizable solution to the challenge of increased fidelity, effectively balancing data diversity with diagnostic precision.

\section{Method}
\label{sec:method}

We developed a multi-encoder framework that processes augmentation-specific inputs in parallel and fuses their features before decoding, enabling robust, augmentation-aware representation learning. Based on this design, we implemented five model configurations to evaluate the contributions of different augmentation and fusion strategies. The first is our proposed model, the multi-encoder builder block (BD), which incorporates a controller network to dynamically weight augmentation-specific features for hierarchical fusion. The second configuration, multi-encoder fusion layer (FL), uses independent encoder-decoder pathways for each augmentation, with outputs averaged volumetrically. The third, multi-encoder concatenation (CC), employs parallel encoder branches to process distinct augmentations, with features fused via channel-wise concatenation. The fourth model uses traditional augmentation (TA), applying standard spatial and geometric transformations during preprocessing in a single-stream network. Finally, the baseline model, with no augmentation (NA), is trained on raw, unaltered input volumes without any augmentation. All models were trained on 204 paired CT–MRI volumes from the OASIS-3 dataset~\cite{lamontagne2019oasis}, validated on 51 additional scans, and evaluated under identical protocols to ensure fair comparison.

\subsection{Model Architectures}

All models employed a U-Net-style backbone with Refined Residual Blocks (RRBs), as described in Section~\ref{sec:method}. Each RRB has two 3x3x3 convolutional layers with ReLU activation. Additive skip connections were used to link the first convolution, which was followed by a 20 \% dropout. The encoder used two downsampling stages with maxpooling to reduce spatial resolution from 128³ to 32³ while increasing feature channels from 64 to 256. The decoder counterpart employed consecutive upsampling layers that integrated upSampling3D operations with transposed convolutions, then enhanced by RRBs to sharpen features. The concluding layer employed a 3×3×3 convolution with sigmoid activation to regenerate the output.

Three key innovations distinguish the proposed MEAL framework from conventional U-Net designs. First, four streams of auto-augmented inputs (flips, rotations, crops, and intensity variations) are processed in parallel by a multi-encoder head, with each augmentation variant assigned to a dedicated encoder branch. Second, augmentation-aware feature fusion is achieved through three distinct strategies: direct concatenation, a dedicated feature fusion layer, and a builder block component, each designed to effectively integrate cross-augmentation representation. Third, a single decoder utilizes the fused multi-augmentation features to reconstruct the output, preserving task-relevant spatial dependencies while maintaining computational efficiency. 

\subsubsection{No-augmentation Learning Method}

The NA and TA methods used a single encoder–decoder pathway, with TA incorporating real-time data augmentation during loading. In contrast, the CC model utilized four parallel encoders, each processing a distinct type of augmented input: flipping, rotation, cropping, or intensity variation. These inputs were passed through identical but non-shared residual encoders, and the resulting feature maps were fused via simple channel-wise concatenation.

\subsubsection{Multi-stream with fusion layer}

The fusion-layer-based multi-stream model (FL) processes four pre-augmented input streams through independent encoder networks. Each encoder $f_{\theta_{k}}$ employs a residual architecture defined by refined residual blocks $R_f$ as shown 

\begin{equation}
    R_f(\mathbf{X})  = \mathbf{X} +\rm Dropout(\rm Conv3D(Conv3D(\mathbf{X})))
\label{eqn:ResidualBlock}
\end{equation}

\noindent with $\mathbf{X}$ the input and $f$ the filter size. The four encoded feature maps $\{h_k\}_{k=1}^{4}$ are fused using parametric feature fusion layer:

\begin{equation}
\mathbf{F}_{\text{FL}} = C_{\text{fuse}} \left( \bigoplus_{k=1}^{4} h_k \right), \quad C_{\text{fuse}} \in \mathbb{R}^{1 \times 1 \times 1 \times 128}
\label{eqn:FL}
\end{equation}

\noindent where $\bigoplus$ denotes channel-wise concatenation and $C_{\rm fuse}$ is a learned 1×1×1 convolutional bottleneck. The innovation lies in its explicit cross-streams to learn the spatial correlations between augmentation types while reducing dimensionality. However, this approach requires separate encoders (quadrupling parameters) and treats augmentations as static, non-adaptive inputs, limiting its capacity to model transformation hierarchies.

\subsubsection{Concatenative Fusion}

The concatenative-fusion-based (CC) model adopts a simpler fusion strategy, processing the same four pre-augmented streams through identical (but non-shared) residual encoders $f_{{\theta}^\prime}$. Features are combined via naïve concatenation,

\begin{equation}
\mathbf{F}_{\text{cc}} = \bigoplus_{k=1}^{4} f_{\theta_k'}(\mathbf{X}_k)
\label{eqn:CC}
\end{equation}

\noindent resulting to a high-dimensional feature space (dim(F)  = 4 ×256). While this approach maximizes feature retention by avoiding compression, it introduces two key limitations: (1) parameter inefficiency, due to non-shared encoder weights across streams, and (2) redundant feature proliferation, which burdens the decoder's capacity to disentangle meaningful signals. Additionally, the absence of cross-stream interactions (e.g., attention or projection mechanisms) increases the risk of overfitting to augmentation-specific artifacts. Its main advantage lies in architectural simplicity, resulting in computationally lightweight inference, albeit at the cost of higher memory usage during training.

\subsubsection{Dynamic controller-Driven Fusion}

This design introduces three paradigm-shifting innovations. First, it replaces static, pre-augmented inputs with learnable, differentiable augmentation modules $\{A_k\}_{k=1}^4$ (e.g., flips, rotations etc.), which are integrated directly into the computational graph. A single, shared encoder $f_{\theta}$ processes all augmented variants such that

\begin{equation}
\mathbf{F}_{\text{BD}} = \sum_{k=1}^{4} \alpha_k f_{\theta}(A_k(\mathbf{X}))
\label{eqn:BD}
\end{equation}

\noindent Second, a controller network computes dynamic attention weights $\alpha_k$ using

\begin{equation}
\alpha_k = \rm{softmax} \left( \mathbf{w}^\top \mathrm{ReLU}(\mathbf{W} \cdot \mathrm{GAP}(h_{\it k})) \right)
\label{eqn:alpha}
\end{equation}

\noindent where GAP is the global average pooling, $\mathbf{W} \in \mathbb{R}^{d \times 256}$, $\mathbf{w} \in \mathbb{R}^{d}$ are learned projections, and $h_k$ the feature map (or feature representation) from the $k^{th}$ encoder–decoder pathway or stream. Third, the decoder $\Gamma_\phi$ reconstruct spatial features through residual transpose convolutions so that 
\begin{equation}
\hat{\mathbf{F}} = \Gamma_\phi(\mathbf{F_{\rm BD}}) = \mathcal{U} \circ R_{128} \circ \mathcal{U} \circ R_{64}(\mathbf{F_{\rm BD}})
\label{eqn:ResidualTranspose}
\end{equation}

\noindent where $\mathcal{U}$  denotes upsampling layer. This architecture reduces parameters by 75\% compared to other methods, while enabling augmentation-aware learning. The controller suppresses irrelevant transformations (low $\alpha_k$) and amplifies informative ones. Crucially, the end-to-end differentiability of $A_k$ allows joint optimization of augmentation parameters and feature weights, a capability absent in other competing methods.

\begin{figure}[htbp]
 \centering
  \includegraphics[width=1.0\textwidth]{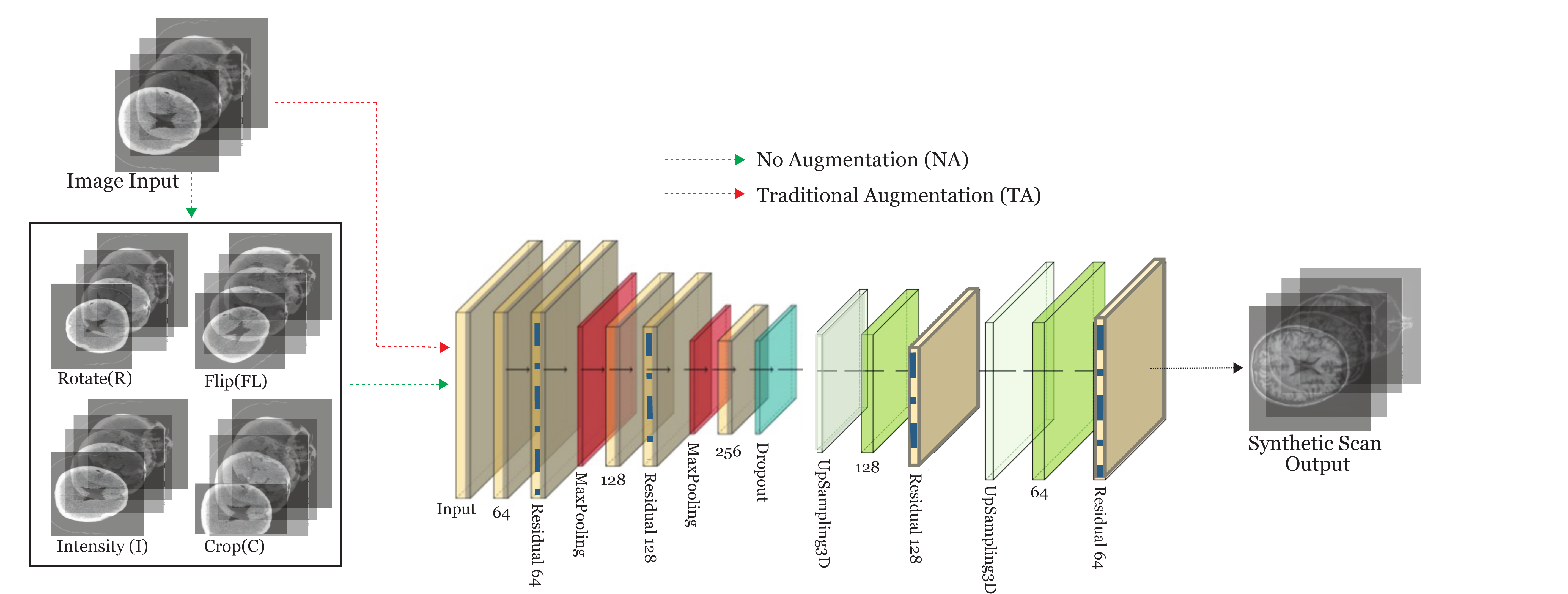}
 \caption{Model architecture for the model having no-augmentation and traditional augmentation.}
\label{fig:MB}
 \end{figure}

\begin{figure}[htbp]
 \centering
  \includegraphics[width=1.0\textwidth]{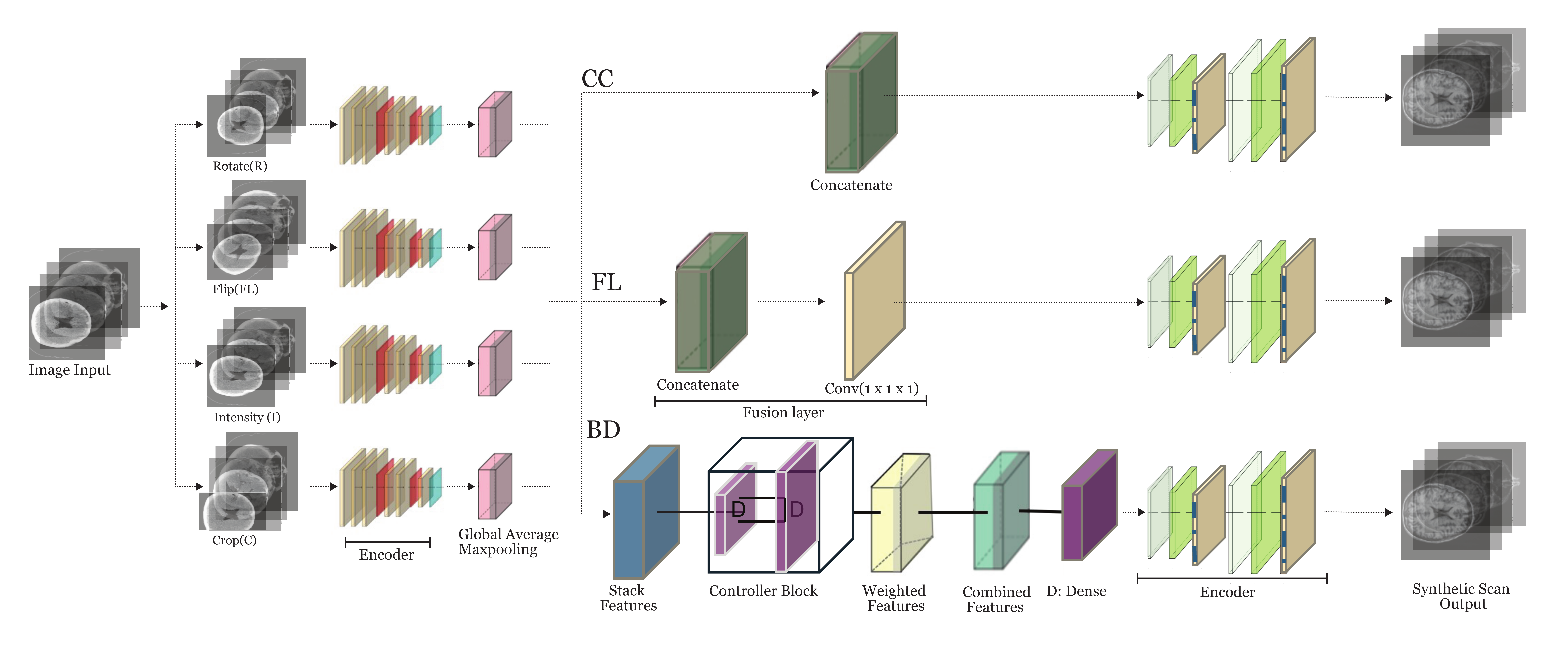}
 \caption{Model architecture for Multi-Stream with a Build Controller method (BD), Fusion layer (FL) and Concatenation (CC).}
\label{fig:ML}
 \end{figure}
 \FloatBarrier
 
% \clearpage
\subsection{Data Preprocessing and Augmentation}

Three-dimensional MRI volumes in NIfTI format were preprocessed by resampling to 128×128×64 voxels using trilinear interpolation, followed by intensity normalization to the range \([0, 1]\)~\cite{shinohara2014statistical}. To enhance soft-tissue contrast, a Hounsfield unit windowing operation was applied (window level = 40, width = 80)~\cite{case2008fundamentals}. To improve model generalization, real-time 3D data augmentations were employed, including random flips along the sagittal and coronal planes ($50\%$ probability), 90° rotations in orthogonal planes, center cropping to 128×128×64, and intensity variations (±$10\%$ brightness/contrast). These augmentations were implemented through parallelized Tensorflow~\cite{abadi2016tensorflow} data pipelines with prefetching to maximize GPU throughput. The OASIS-3 dataset~\cite{lamontagne2019oasis} provided 255 paired CT and T1-weighted MRI volumes. Each pair underwent rigid spatial registration using ANTs~\cite{tustison2024antsx,tustison2021antsx}, followed by the same preprocessing steps: intensity normalization, trilinear resizing to 128×128×64, and CT windowing (level = 40, width = 80). For augmented model variants (CC, FL, and BD), additional spatial transformations (random flips, ±15° rotations, and resizing from 100³ to 128³) and intensity perturbations (±$10\%$ scaling and shifts) were applied. Volumes were batched as 128×128×64×1 tensors and prefetched to optimize GPU utilization.

\subsubsection{Random Spatial Flip}
This layer applies independent random flips along the height ($h$) and width ($w$) axes with probability $p = 0.5$. For each input tensor $\mathbf{X}$, two Bernoulli trials $b_h, b_w \sim \mathcal{U}(0, 1)$ determine whether flipping occurs:

\begin{equation}
    \mathbf{X}_h =
\begin{cases}
\text{flip}(\mathbf{X}, \text{axis} = 1) & \text{if } b_h > 0.5 \\
\mathbf{X} & \text{otherwise}
\end{cases}
\end{equation}

and
\begin{equation}
    \mathbf{X}_w =
\begin{cases}
\text{flip}(\mathbf{X}_h, \text{axis} = 2) & \text{if } b_w > 0.5 \\
\mathbf{X}_h & \text{otherwise}
\end{cases}
\end{equation}
The \texttt{flip()} function reverses the tensor along the specified axis, introducing reflection symmetry while preserving dimensionality.

\subsubsection{Random Orthogonal Rotation}

The input is rotated by $k \times 90^\circ$, where $k \sim \mathcal{U}\{0, 1, 2, 3\}$. To handle 3D inputs, the tensor is reshaped to $\mathbf{X}_{\text{reshaped}} \in \mathbb{R}^{B \cdot D \times H \times W \times C}$ and rotated as $\mathbf{X}_{\text{rot}} = \text{rot90}(\mathbf{X}_{\text{reshaped}}, k)$. Then, the tensor is restored to its original shape, ensuring invariance to orthogonal planar rotations without distorting depth or channels.

\subsubsection{Center Crop and Resize}

A subvolume of size $(H_c, W_c, D_c)$ is cropped from the center of the input and resized back to $(H, W, D)$ using bilinear interpolation. The crop offsets are computed as

\begin{equation}
    \Delta h = \left\lfloor \frac{H - H_c}{2} \right\rfloor, \quad 
\Delta w = \left\lfloor \frac{W - W_c}{2} \right\rfloor, \quad 
\Delta d = \left\lfloor \frac{D - D_c}{2} \right\rfloor
\end{equation}

\noindent with cropped tensor denoted as $\mathbf{X}_{\text{resized}} = \text{resize}\left( \mathbf{X}_{\text{crop}}, (H, W) \right)$. This simulates varying fields of view while maintaining spatial dimensions.

\subsubsection{Intensity Perturbation}

To simulate illumination variability, global intensity transformations are applied through brightness and contrast adjustments. The brightness is modified by adding a random scalar shift $\delta$ to the input tensor as shown

\begin{equation}
    \mathbf{X}' = \mathbf{X} + \delta, \quad \delta \sim \mathcal{U}(-0.1, 0.1)
\end{equation}

\noindent The contrast is adjusted by scaling the brightness-modified tensor relative to its mean intensity:

\begin{equation}
    \mathbf{X}'' = \alpha \mathbf{X}' + (1 - \alpha) \mu_{\mathbf{X}}, \quad \alpha \sim \mathcal{U}(0.9, 1.1)
\end{equation}

\noindent where $\mu_{\mathbf{X}}$ denotes the mean intensity of the original tensor $\mathbf{X}$.

\subsection{Training Protocol and Implementation}

All models were trained for 300 epochs using the Adam optimizer with an initial learning rate of 1e-4. Then, we define loss function $\mathcal L$ as
\begin{equation}
    \mathcal L = \mathcal L_{1} + 0.8 (1-\rm SSIM)
\end{equation}
\noindent with SSIM denotes the structural fidelity in the reconstructed volumes, and $\mathcal L_{1}$ the mean square error. The learning rate was reduced by $50\%$, if the validation loss plateaued for more than five consecutive epochs. Training was conducted on NVIDIA A6000 GPUs using Tensorflow 2.10~\cite{abadi2016tensorflow}. Due to memory constraints, each batch contained a single 3D volume. Models were trained with checkpointing based on the lowest validation loss and batch-wise logging to enable consistent monitoring. Reproducibility was ensured by fixing random seeds across all libraries. To isolate the effect of the fusion strategy, augmentation parameters were standardized across all model variants.

\subsection{Statistical Analysis}

Model performance was evaluated using 3D PSNR and SSIM metrics computed using tensorflow-mri~\cite{montalt2022tensorflow} across the full 128$^3$-voxel volumes. Structural accuracy was further assessed using Dice similarity coefficients~\cite{raina2023tackling} between reconstructed and ground-truth MRI segmentations of white and gray matter. To evaluate differences in image reconstruction quality across methods, we analyzed two quantitative metrics, such as PSNR and SSIM. Pairwise statistical comparisons were conducted against a designated reference methods. For each comparison, metric differences were computed and assessed for normality using the Shapiro–Wilk test~\cite{razali2011power}. Based on the normality results, either a paired t-test (for normally distributed differences) or a Wilcoxon signed-rank test~\cite{rosner2006wilcoxon} (for non-normal differences) was applied.

To determine whether significant global differences existed among multiple methods, normality was again assessed across all groups using the Shapiro–Wilk test~\cite{razali2011power}. If normality was satisfied, a one-way ANOVA was employed; otherwise, a Kruskal–Wallis H test was used~\cite{ostertagova2014methodology}. In cases where the global test indicated significance, Dunn’s post-hoc test with Bonferroni correction was performed to identify specific group differences~\cite{ruxton2008time}. Statistical significance was defined as a p-value $< \alpha$ (0.01), where $\alpha$ denotes the significance level. Methods were further ranked based on mean metric scores to provide comparative insight. All statistical analyses were performed using the scipy.stats~\cite{seabold2010statsmodels,virtanen2020scipy} and scikit-posthocs~\cite{terpilowski2019scikit} libraries in Python. Qualitative evaluation included voxel-wise error maps and attention heatmaps for the BD model, visualized using consistent windowing parameters (window level = 40, window width = 80).

\subsection{Brain Tissue Segmentation}
We performed brain tissue segmentation using a dual-path pipeline developed using ANTs library~\cite{tustison2021antsx,tustison2024antsx} in Python. The workflow was separated for synthetic and ground-truth (GT) T1w MRI to account for differences in noise and intensity characteristics. For both paths, preprocessing consisted of isotropic resampling to 1 mm³ and intensity normalization to the [0–1] range. Adaptive skull stripping~\cite{laha2018skull} was applied using a multi-stage approach for synthetic T1 scans (following ANTs extraction $\rightarrow$ Otsu thresholding $\rightarrow$ hybrid ANTs–Otsu)~\cite{sang2024improved}, while GT scans used a conservative ANTs extraction with minimal refinement. Moreover, the N4 bias field correction~\cite{kanakaraj2024deepn4} was then applied to all skull-stripped images. Tissue segmentation was performed using ANTs Atropos~\cite{tustison2021antsx,tustison2024antsx}, an expectation–maximization algorithm with spatial regularization. Parameters were optimized per image type: synthetic scans used robust settings (m = 0.2; c = [3,0]), whereas GT scans automatically selected one of three parameter sets (conservative: m=0.1, 10 iterations; moderate: m=0.2, 5 iterations; aggressive: m=0.3, 3 iterations) based on physiologically expected tissue proportions (cerebrospinal fluid (CSF): 5–30\%, gray matter (GM): 30–60\%, white matter (WM): 20–50\%). Both workflows generated three-class segmentations (CSF, GM, WM) and included fallback mechanisms, such as percentile-based thresholding~\cite{jardim2023image}, to address possible segmentation failures. Volumetric metrics were derived from the segmentation labels using voxel counting, converted from $\mathrm{mm}^3$ to $\mathrm{mL}$, and the brain parenchymal fraction $\mathrm{BPF}$ was computed as $\mathrm{BPF} = \frac{\mathrm{GM} + \mathrm{WM}}{\mathrm{ICV}} \times 100\%$, where $\mathrm{ICV}$ is intracranial volume. Segmentation performance was evaluated using Dice similarity coefficients~\cite{cardenas2018deep}, and intensity-difference histograms.

\subsection{Trainable Parameters and Computational Resources}

The computational complexity of each reconstruction method was assessed by comparing the total number of trainable parameters. The BD model had the highest complexity, with 273,265,538 trainable parameters, followed by the CC and FL models with 13,725,953 and 10,760,577 parameters respectively. The NA/TA model was the most lightweight, comprising only 4,428,545 trainable parameters. All models reported zero non-trainable parameters, indicating that the entire model architecture was optimized during training. In addition, model training and evaluation were conducted on a high-performance computing system running Ubuntu 22.04 with an 18-core (36-thread) Intel x86$\_$64 CPU, 188 GB of RAM, and four NVIDIA RTX A6000 GPUs, each with approximately 48 GB of VRAM. During experimentation, GPU utilization, memory allocation, and temperature were actively monitored to ensure efficient and stable training.

%%%%
% \clearpage
\section{Results}

\subsection{Feature Learning without Augmentation}

% \subsubsection{Quantitative Comparison}

\begin{figure}[htbp] %h!
 \centering
  \includegraphics[width=1.0\textwidth]{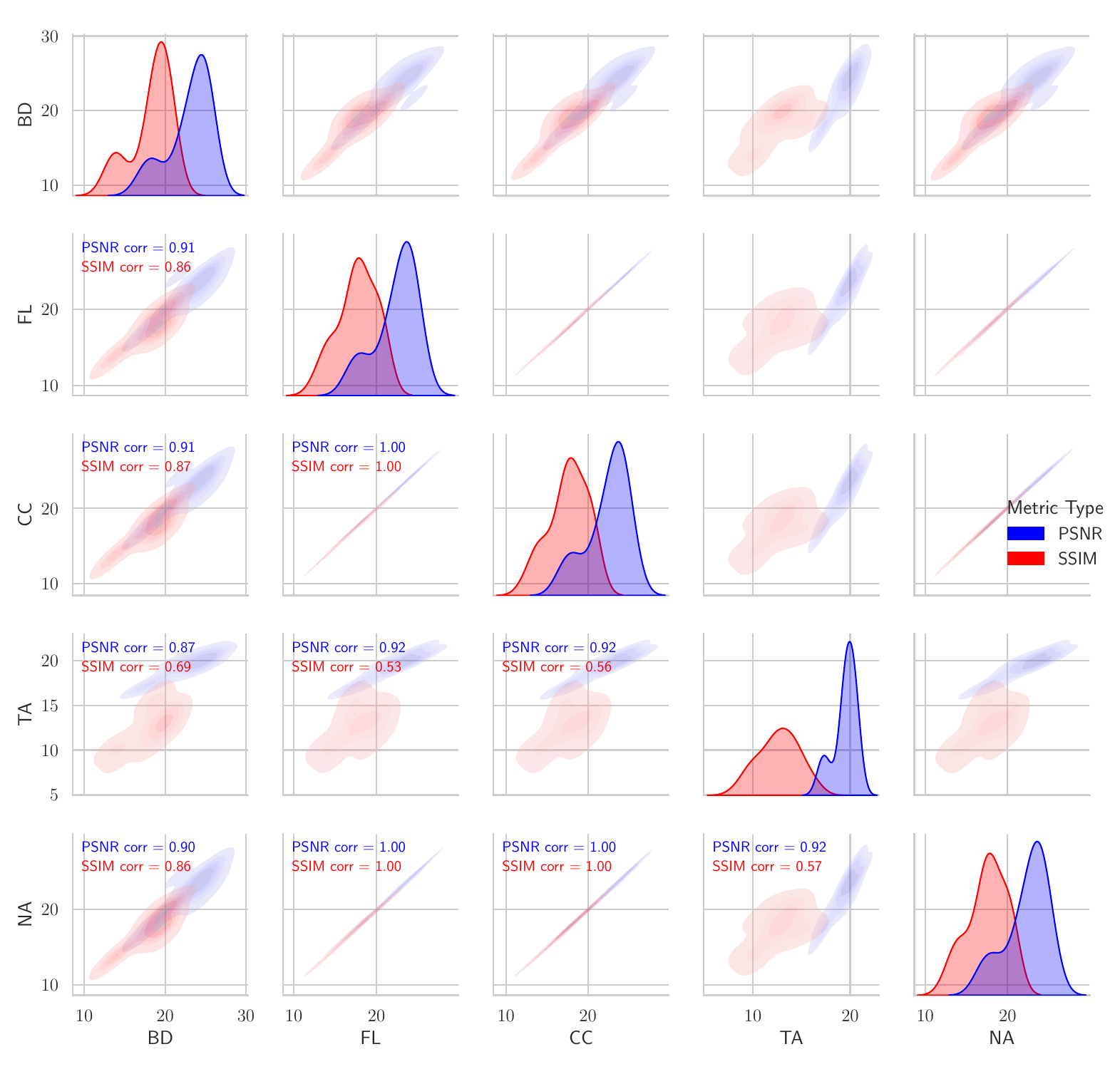}
 \caption{Pairwise comparison of PSNR and SSIM across reconstruction methods under no-augmentation with unseen testing data.}
\label{fig:NoAUg Unseen Testing Data}
 \end{figure}

Figures~\ref{fig:NoAUg Unseen Testing Data} and S12 show pairwise comparisons of PSNR and SSIM across five reconstruction methods (BD, FL, CC, TA, and NA) under no-augmentation conditions, with Figure~\ref{fig:NoAUg Unseen Testing Data} evaluating unseen test data and Figure S12 explaining the predefined testing set. In Figure~\ref{fig:NoAUg Unseen Testing Data}, both PSNR and SSIM show strong correlations across methods (mostly exceeding 0.85), with overlapping distributions and consistent pairwise trends, indicating stable reconstruction performance on previously unseen inputs. However, Figure S12 reveals a notable decline in metric correlations (which dropped from 0.58 to 0.10 in some cases) and an increased divergence between the PSNR and SSIM distributions. This discrepancy shows that even when models generalize well to new data, performance variability can still occur within the test set. 

\clearpage

\begin{figure}[htbp] %h!
 \centering
  \includegraphics[width=0.85\textwidth]{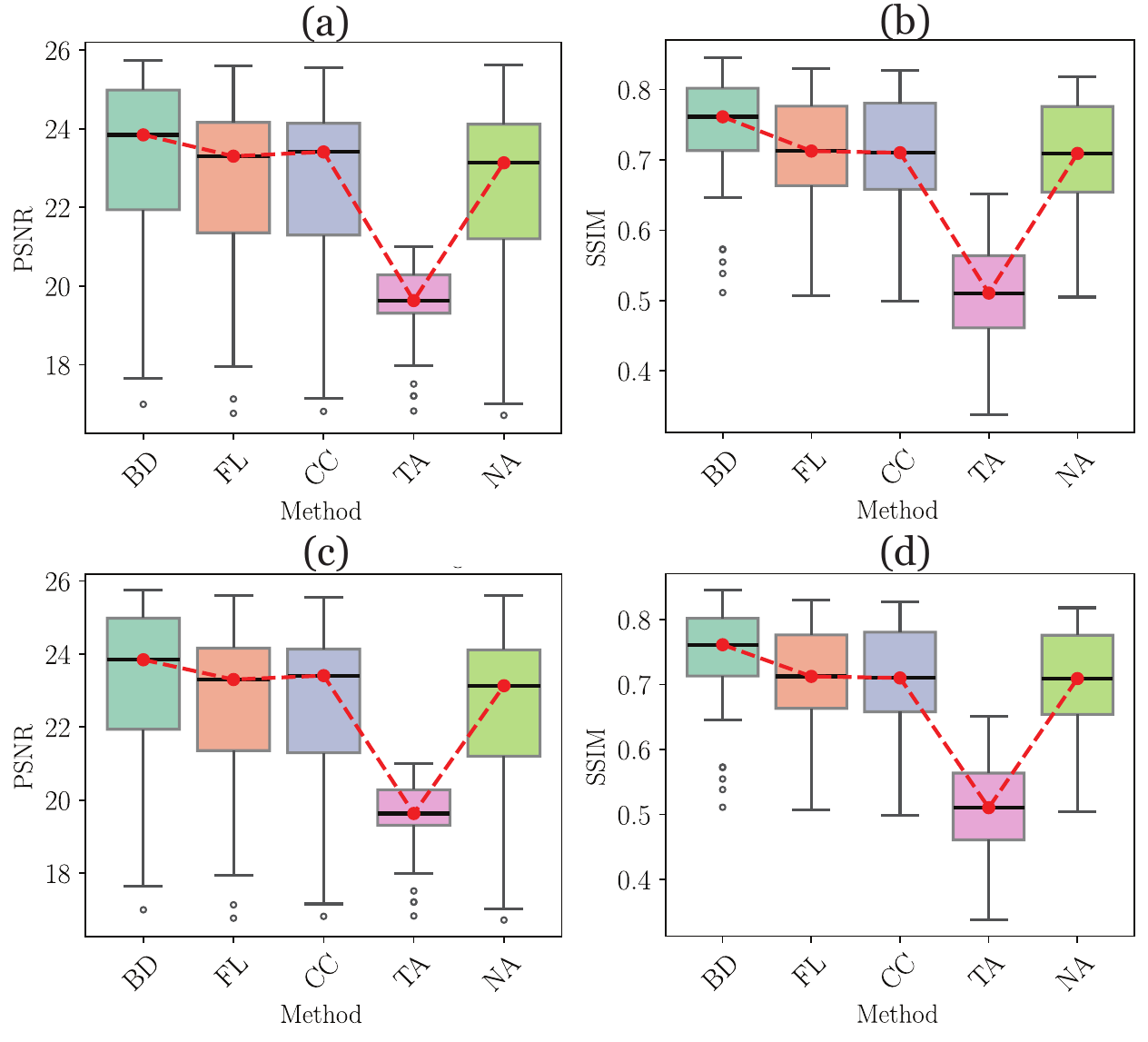}
 \caption{Boxplots showing PSNR and SSIM distributions for five methods (BD, FL, CC, TA, NA) evaluated on unseen (panels (a) and (b)) and predefined test (panels (c) and (d)) data under no-augmentation.}
\label{fig:noaug_boxplot}
 \end{figure}

\noindent Furthermore, Figure~\ref{fig:noaug_boxplot} confirms consistent trends in PSNR and SSIM across unseen- (panels (a) and (b), Figure~\ref{fig:noaug_boxplot}) and predefined-test (panels (c) and (d), Figure~\ref{fig:noaug_boxplot}) data, with BD achieving the highest scores and the other methods showing similar and overlapping performance.

\begin{figure}[htbp] %h
 \centering
  \includegraphics[width=0.9\textwidth]{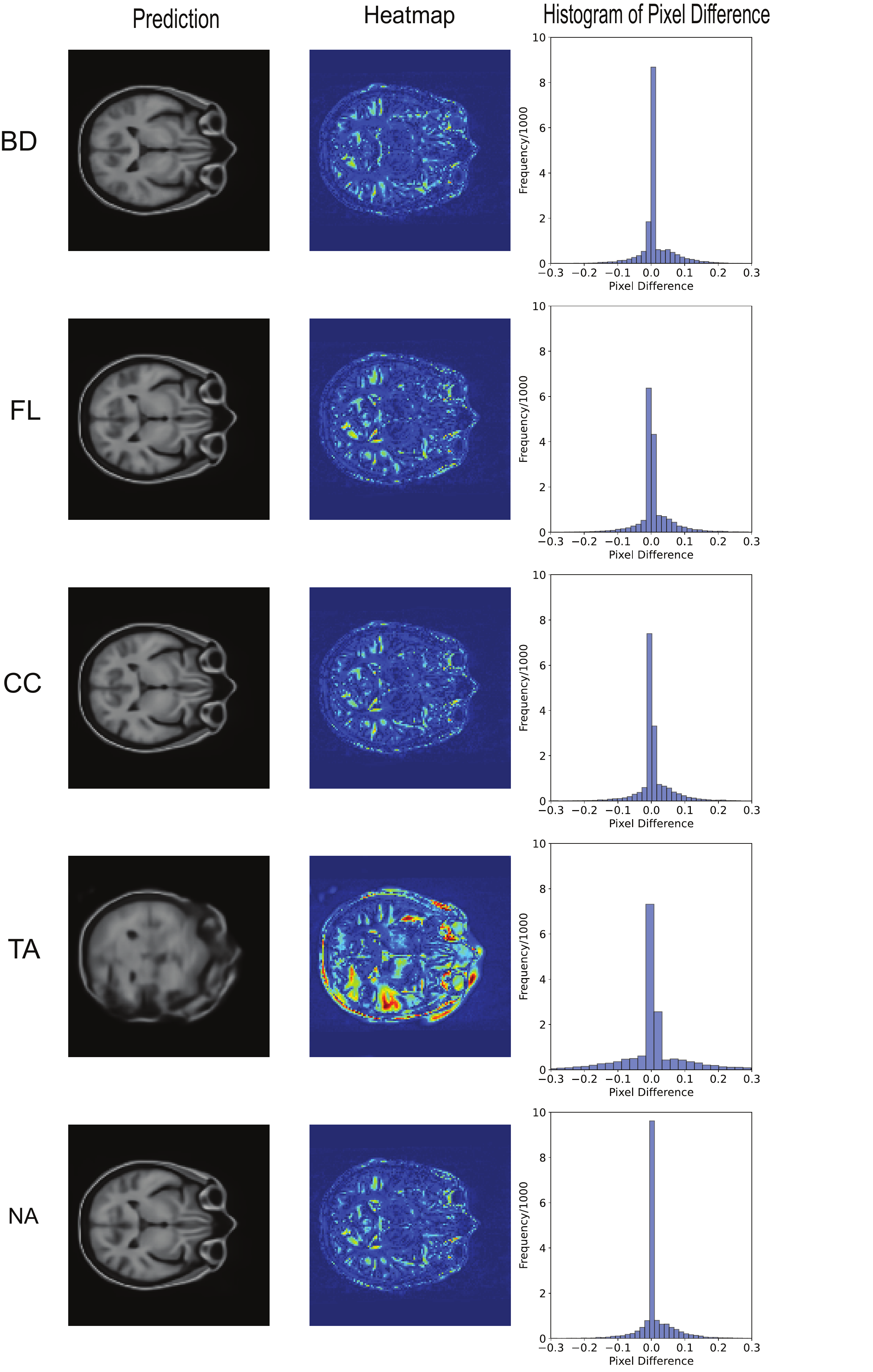}
 \caption{Pixel-wise reconstruction error analysis under no-augmentation (unseen-test data) using heatmaps and histograms.}
\label{fig:unseen_noaug}
 \end{figure}

Figure~\ref{fig:unseen_noaug} presents a detailed pixel-wise error analysis for the candidate reconstruction methods (BD, FL, CC, TA, NA), showing the predicted images, error heatmaps, and histograms of pixel differences. BD exhibits the lowest overall error, with minimal visible artifacts in the heatmap and a tightly concentrated histogram centered near zero, indicating high reconstruction fidelity. NA and CC also display favorable error distributions, with relatively uniform heatmaps and moderately concentrated histograms, suggesting effective preservation of structural details. Although CC does not show significant improvements in PSNR, its balanced error distribution and reduced structural artifacts highlights its reliability for structural preservation, consistent with earlier SSIM-based observations. 

However, FL shows slightly higher localized errors, while TA exhibits the highest error concentration, with prominent red and yellow regions in the heatmap and a broader histogram spread, reflecting substantial pixel-level deviations (middle panel, Figure~\ref{fig:unseen_noaug}). These findings align with overall performance trends and further support NA and CC as the most robust methods for preserving image quality and structural integrity, particularly in scenarios without data augmentation.

 \FloatBarrier
\clearpage

\clearpage

\begin{table}[htbp] %h!
\centering
\caption{Summary of statistical analysis across methods under no-augmentation.}
\scriptsize
\resizebox{0.95\textwidth}{!}{%
\begin{tabular}{llllll}
\toprule
\textbf{Section} & \textbf{Comparison} & \textbf{Metric} & \textbf{Test used} & \textbf{p-value} & \textbf{Significant?} \\
\midrule

\multicolumn{6}{l}{\textbf{Unseen-test data}} \\
\midrule
Group-wise  & All Methods & PSNR & Kruskal-Wallis & \textbf{8.89e-07} & Yes \\
Group-wise  & All Methods & SSIM & Kruskal-Wallis & \textbf{4.19e-11} & Yes \\
\midrule
Dunn's  & BD vs TA & PSNR & Dunn (Bonferroni) & \textbf{8.45e-07} & Yes \\
& CC vs TA & PSNR & Dunn (Bonferroni) & \textbf{3.03e-04} & Yes \\
& FL vs TA & PSNR & Dunn (Bonferroni) & \textbf{3.40e-04} & Yes \\
& NA vs TA & PSNR & Dunn (Bonferroni) & \textbf{6.25e-04} & Yes \\
Dunn's  & BD vs TA & SSIM & Dunn (Bonferroni) & \textbf{3.16e-10} & Yes \\
& CC vs TA & SSIM & Dunn (Bonferroni) & \textbf{6.66e-07} & Yes \\
& FL vs TA & SSIM & Dunn (Bonferroni) & \textbf{3.11e-07} & Yes \\
& NA vs TA & SSIM & Dunn (Bonferroni) & \textbf{1.59e-06} & Yes \\
\midrule

\textbf{Condition} & \textbf{Method} & \textbf{Metric} & \textbf{PSNR} & \textbf{SSIM} & -- \\
\midrule
\multirow{5}{*}{Unseen} 
& BD & -- & \textbf{23.03} & \textbf{0.733} & \\
& FL & -- & 22.38 & 0.707 & \\
& CC & -- & 22.41 & 0.704 & \\
& NA & -- & 22.32 & 0.701 & \\
& TA & -- & 19.48 & 0.506 & \\
\midrule

\multicolumn{6}{l}{\textbf{Predefined-test data}} \\
\midrule
Group-wise  & All Methods & PSNR & Kruskal-Wallis & \textbf{2.29e-06} & Yes \\
Group-wise  & All Methods & SSIM & Kruskal-Wallis & \textbf{6.68e-13} & Yes \\
\midrule
Dunn's  & BD vs TA & PSNR & Dunn (Bonferroni) & \textbf{5.12e-06} & Yes \\
& CC vs TA & PSNR & Dunn (Bonferroni) & \textbf{1.23e-03} & Yes \\
& FL vs TA & PSNR & Dunn (Bonferroni) & \textbf{1.45e-03} & Yes \\
& NA vs TA & PSNR & Dunn (Bonferroni) & \textbf{4.52e-04} & Yes \\
Dunn's  & BD vs TA & SSIM & Dunn (Bonferroni) & \textbf{1.92e-09} & Yes \\
& CC vs TA & SSIM & Dunn (Bonferroni) & \textbf{4.33e-06} & Yes \\
& FL vs TA & SSIM & Dunn (Bonferroni) & \textbf{2.10e-06} & Yes \\
& NA vs TA & SSIM & Dunn (Bonferroni) & \textbf{9.24e-07} & Yes \\
\midrule

\textbf{Condition} & \textbf{Method} & \textbf{Metric} & \textbf{PSNR} & \textbf{SSIM} & -- \\
\midrule
\multirow{5}{*}{Validation} 
& BD & -- & \textbf{24.12} & \textbf{0.745} & \\
& FL & -- & 22.84 & 0.712 & \\
& CC & -- & 22.92 & 0.715 & \\
& NA & -- & 22.88 & 0.710 & \\
& TA & -- & 20.18 & 0.512 & \\
\bottomrule
\end{tabular}%
}
\label{table:combined_noaug_results}
\end{table}

Table \ref{table:combined_noaug_results} presents a statistical summary for reconstruction performance under no-augmentation for both unseen and predefined-test data. Across both datasets, the Kruskal-Wallis test identified significant differences in PSNR and SSIM among all methods (p-value $<$ $\alpha$ (0.01)). Post-hoc Dunn’s tests with Bonferroni correction consistently showed that BD significantly outperformed TA in both PSNR and SSIM on unseen- (p-value $<$ $\alpha$(0.01) and predefined-test data (p-value $<$ $\alpha$(0.01)). Other methods (CC, FL, NA) also demonstrated statistically significant improvements over TA, though to a lesser extent. BD achieved the highest average PSNR (23.03 dB and 24.12 dB) and SSIM (0.733 and 0.745) on unseen- and predefined-test data, respectively, affirming its superior reconstruction quality. Conversely, TA consistently exhibited the lowest performance across both metrics and datasets. These findings are consistent with the heatmap and histogram visualizations in Figure \ref{fig:unseen_noaug}, further corroborating that BD pr-ovides the highest pixel-wise fidelity under no-augmentation.

\subsection{Augmentation Specific Feature Learning}
In this section, we investigate augmentation learning by evaluating the effect of different transformation strategies on model generalization and robustness, assessing reconstruction performance under various augmentation conditions on both unseen and predefined-test data.

\subsubsection{Comparative Analysis of Reconstruction Metrics}
\begin{figure}[htbp] %h!
 \centering
  \includegraphics[width=1.0\textwidth]{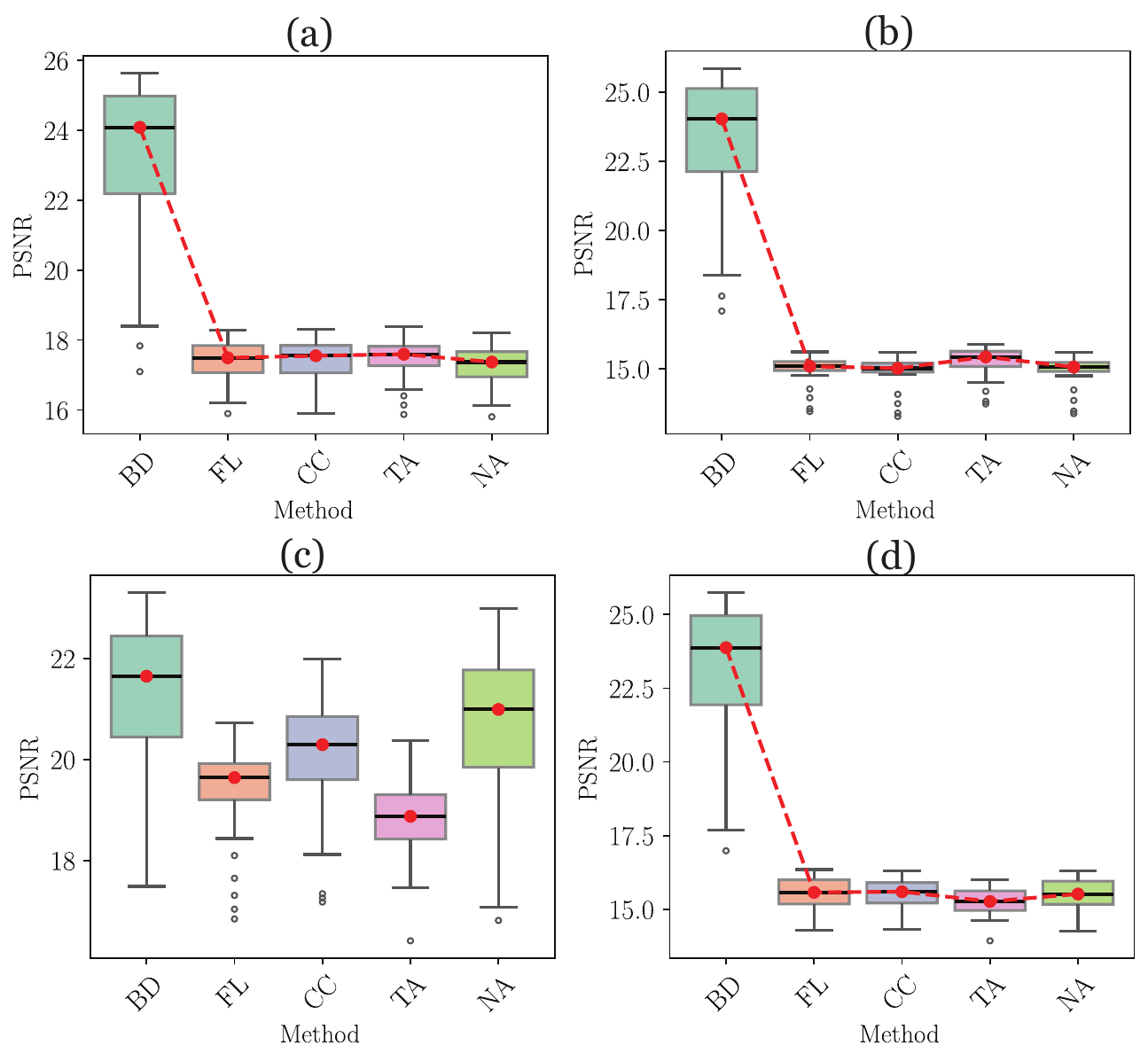}
 \caption{Boxplot showing PSNR distributions for five methods (BD, FL, CC, TA, NA) evaluated on unseen test data under (a) rotation (b) crop (c) flip and (d) intensity augmentation.}
\label{fig:AugPixelWise}
 \end{figure}

The boxplots in Figures~\ref{fig:AugPixelWise}, S5, S13, and S14 summarize the distribution of PSNR and SSIM scores across five reconstruction methods (BD, FL, CC, TA, and NA) evaluated on both validation datasets. Consistently across all conditions, BD outperforms the other methods, achieving the highest median PSNR and SSIM values. TA, on the other hand, demonstrates the poorest performance across all scenarios. These trends confirm the robustness of BD across diverse augmentation types while highlighting the vulnerability of TA to performance degradation. Furthermore, Figures S1-S4 show the pairwise correlation between the distributions of PSNR and SSIM across candidate reconstruction methods (BD, FL, CC, TA, NA) under various augmentation and validation data (Figures S8-S11). BD consistently demonstrates the highest PSNR and SSIM values with tightly clustered distributions, reflecting stable and high-quality reconstructions. Conversely, other methods show wider distributions and lower correlations, especially TA, which frequently exhibits weak SSIM correlations, indicating inconsistent structural preservation. Variability in metric correlation patterns is more pronounced under crop and flip augmentations, where several methods (e.g., FL, TA) display low or near-zero SSIM correlations despite moderate PSNR correlations. These findings confirm BD’s metric stability and reveal that PSNR and SSIM respond differently across methods and augmentations, highlighting the need to consider both when evaluating reconstruction performance.
\clearpage

Here, the statistical analysis of the reconstruction performance, under various augmentation strategies in both validation data sets is displayed in Tables \ref{table:unseen_data_summary} and S1. Here, Kruskal-Wallis tests revealed significant differences in PSNR and SSIM across all methods (p-value $<$ $\alpha$ (0.01)) under every augmentation type. As already shown in Section, post-hoc Dunn’s tests with Bonferroni correction also consistently identified BD as significantly superior to all competing methods in both PSNR and SSIM, particularly outperforming TA with highly significant p-values (p-value $<$ $\alpha$ (0.01) across all augmentations). This demonstrates the robustness of BD to geometric and photometric transformations. Conversely, TA consistently showed the lowest performance. Notably, BD maintained PSNR scores above 23 dB and SSIM above 0.72 across most augmentation types, confirming its ability to generalize well under varying perturbations.

\begin{table}[H] %h!
\centering
\caption{Unseen-test data: Summary of statistical analysis across methods for rotate, crop, intensity, and flip augmentations.}
\scriptsize
% \resizebox{\textwidth}{!}{%
\resizebox{0.85\textwidth}{!}{%
\begin{tabular}{llllll}
\toprule
\textbf{Section} & \textbf{Comparison} & \textbf{Metric} & \textbf{Test Used} & \textbf{p-value} & \textbf{Significant?} \\
\midrule

\multicolumn{6}{l}{\textbf{Group-wise Tests}} \\
\midrule
Group-wise (Rotate) & All Methods & PSNR & Kruskal-Wallis & \textbf{0.0000} & Yes \\
Group-wise (Rotate) & All Methods & SSIM & Kruskal-Wallis & \textbf{0.0000} & Yes \\
Group-wise (Crop)   & All Methods & PSNR & Kruskal-Wallis & \textbf{0.0000} & Yes \\
Group-wise (Crop)   & All Methods & SSIM & Kruskal-Wallis & \textbf{0.0000} & Yes \\
Group-wise (Intensity) & All Methods & PSNR & Kruskal-Wallis & \textbf{3.05e-09} & Yes \\
Group-wise (Intensity) & All Methods & SSIM & Kruskal-Wallis & \textbf{8.28e-09} & Yes \\
Group-wise (Flip) & All Methods & PSNR & Kruskal-Wallis & \textbf{0.0000} & Yes \\
\midrule

\multicolumn{6}{l}{\textbf{Dunn's Test (PSNR and SSIM)}} \\
\midrule
Dunn's (Rotate) & BD vs CC & PSNR & Dunn (Bonferroni) & \textbf{5.38e-08} & Yes \\
& BD vs FL & PSNR & Dunn (Bonferroni) & \textbf{4.50e-08} & Yes \\
& BD vs NA & PSNR & Dunn (Bonferroni) & \textbf{1.97e-10} & Yes \\
& BD vs TA & PSNR & Dunn (Bonferroni) & \textbf{1.37e-07} & Yes \\
& CC vs FL & PSNR & Dunn (Bonferroni) & 1.0000 & No \\
& BD vs CC & SSIM & Dunn (Bonferroni) & \textbf{2.47e-08} & Yes \\
& BD vs FL & SSIM & Dunn (Bonferroni) & \textbf{1.45e-07} & Yes \\
& BD vs NA & SSIM & Dunn (Bonferroni) & \textbf{1.16e-08} & Yes \\
& BD vs TA & SSIM & Dunn (Bonferroni) & \textbf{4.86e-15} & Yes \\
Dunn's (Crop) & BD vs CC & PSNR & Dunn (Bonferroni) & \textbf{1.20e-12} & Yes \\
& BD vs FL & PSNR & Dunn (Bonferroni) & \textbf{2.02e-10} & Yes \\
& BD vs NA & PSNR & Dunn (Bonferroni) & \textbf{9.13e-12} & Yes \\
& BD vs TA & PSNR & Dunn (Bonferroni) & \textbf{5.00e-05} & Yes \\
& CC vs FL & PSNR & Dunn (Bonferroni) & 1.0000 & No \\
& BD vs CC & SSIM & Dunn (Bonferroni) & \textbf{2.95e-10} & Yes \\
& BD vs FL & SSIM & Dunn (Bonferroni) & \textbf{2.78e-11} & Yes \\
& BD vs NA & SSIM & Dunn (Bonferroni) & \textbf{4.78e-08} & Yes \\
& BD vs TA & SSIM & Dunn (Bonferroni) & \textbf{2.31e-09} & Yes \\
Dunn's (Intensity) & BD vs FL & PSNR & Dunn (Bonferroni) & \textbf{6.99e-05} & Yes \\
& BD vs TA & PSNR & Dunn (Bonferroni) & \textbf{3.54e-08} & Yes \\
& NA vs FL & PSNR & Dunn (Bonferroni) & \textbf{0.0126} & Yes \\
& NA vs TA & PSNR & Dunn (Bonferroni) & \textbf{3.56e-05} & Yes \\
& BD vs FL & SSIM & Dunn (Bonferroni) & \textbf{1.08e-05} & Yes \\
& BD vs TA & SSIM & Dunn (Bonferroni) & \textbf{1.48e-05} & Yes \\
& FL vs NA & SSIM & Dunn (Bonferroni) & \textbf{0.00096} & Yes \\
Dunn's (Flip) & BD vs CC & PSNR & Dunn (Bonferroni) & \textbf{1.58e-08} & Yes \\
& BD vs FL & PSNR & Dunn (Bonferroni) & \textbf{4.07e-08} & Yes \\
& BD vs NA & PSNR & Dunn (Bonferroni) & \textbf{8.52e-09} & Yes \\
& BD vs TA & PSNR & Dunn (Bonferroni) & \textbf{7.31e-14} & Yes \\
\midrule

\multicolumn{6}{l}{\textbf{Mean Scores (PSNR and SSIM)}} \\
\midrule
\textbf{Augmentation} & \textbf{Method} & \textbf{Metric} & \textbf{PSNR} & \textbf{SSIM} & -- \\
\midrule
\multirow{5}{*}{Rotate} 
& BD & -- & \textbf{23.105} & \textbf{0.721} & \\
& TA & -- & 17.439 & 0.377 & \\
& FL & -- & 17.407 & 0.425 & \\
& CC & -- & 17.397 & 0.422 & \\
& NA & -- & 17.277 & 0.420 & \\
\midrule
\multirow{5}{*}{Crop} 
& BD & -- & \textbf{23.122} & \textbf{0.727} & \\
& TA & -- & 15.256 & 0.260 & \\
& FL & -- & 14.965 & 0.254 & \\
& NA & -- & 14.915 & 0.263 & \\
& CC & -- & 14.883 & 0.256 & \\
\midrule
\multirow{5}{*}{Intensity} 
& BD & -- & \textbf{21.210} & \textbf{0.611} & \\
& NA & -- & 20.590 & 0.582 & \\
& CC & -- & 20.054 & 0.577 & \\
& FL & -- & 19.327 & 0.475 & \\
& TA & -- & 18.824 & 0.474 & \\
\midrule
\multirow{5}{*}{Flip} 
& BD & -- & \textbf{23.021} & \textbf{0.733} & \\
% & BD & -- & \textbf{23.021} & \textbf{0.733} & \\
& FL & -- & 15.564 & 0.339 & \\
& CC & -- & 15.540 & 0.338 & \\
& NA & -- & 15.523 & 0.337 & \\
& TA & -- & 15.248 & 0.308 & \\
\bottomrule
\end{tabular}%
}
\label{table:unseen_data_summary}
\end{table}

\clearpage
\subsection{Robustness to Clinical Variability}

\begin{figure}[htbp] %h
 \centering
  \includegraphics[width=0.9\textwidth]{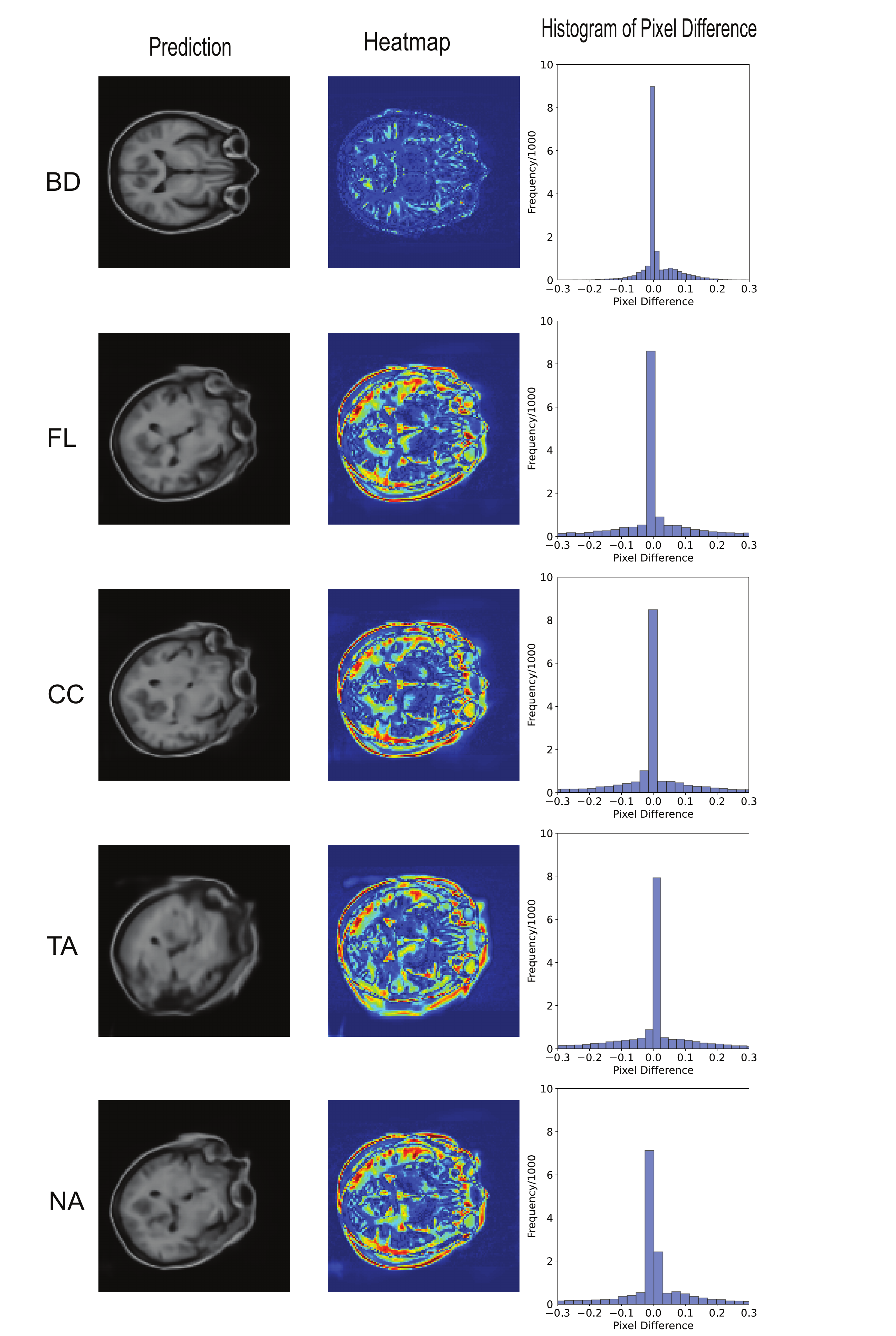}
 \caption{Pixel-wise reconstruction error analysis under rotation augmentation (unseen test data) using heatmaps and histograms.}
\label{fig:Rotation_unseen}
 \end{figure}

\begin{figure}[htbp] %h
 \centering
  \includegraphics[width=0.9\textwidth]{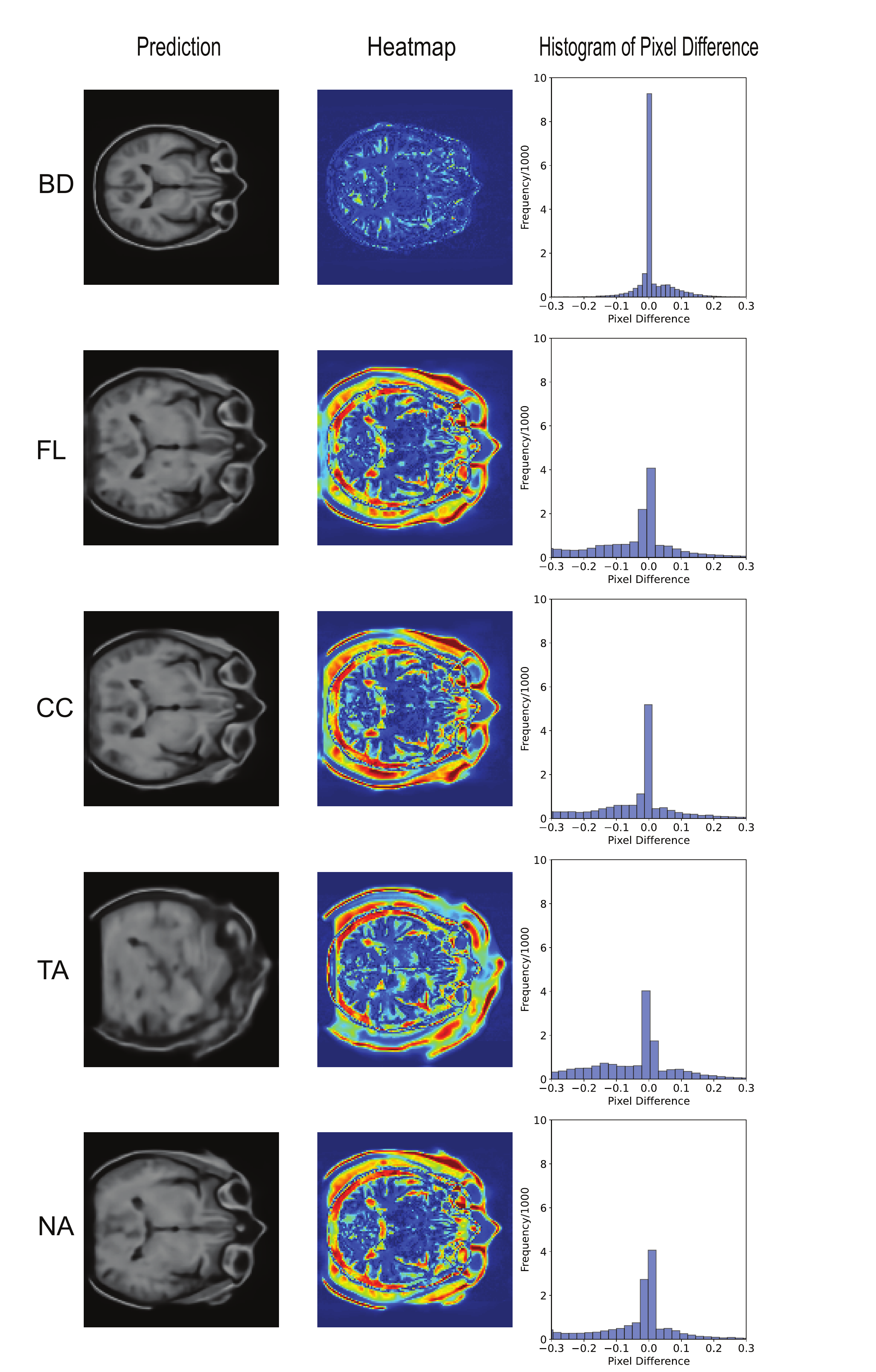}
 \caption{Pixel-wise reconstruction error analysis under crop augmentation (unseen test data) using heatmaps and histograms.}
\label{fig:Crop_unseen}
 \end{figure}

To ensure practical applicability in clinical settings, models must generalize across variations caused by patient positioning, anatomical coverage, and acquisition conditions. This section evaluates robustness under various perturbations. We present pixel-wise reconstruction error analyses for these perturbations, such as rotation (Figure~\ref{fig:Rotation_unseen}), cropping (Figure~\ref{fig:Crop_unseen}), flipping (Figure S6), and intensity variation (Figure S7). These transformations simulate realistic challenges in clinical imaging, including geometric misalignments and scanner-induced intensity inconsistencies, which models must robustly handle in practice.

Across all scenarios, BD consistently achieves the lowest reconstruction errors, evidenced by the dominance of cool blue regions in the heatmaps and narrow, sharply peaked histograms centered near zero pixel difference. For instance, in Figure S6 (flip), BD maintains minimal error across the entire image, while other methods such as TA and FL exhibit widespread spatial artifacts highlighted by extensive red and yellow regions, indicating vulnerability to simple geometric transformations.
Similarly, Figure~\ref{fig:Rotation_unseen} (rotation) reinforces this observation, where BD shows minimal spatial error, whereas TA, NA, and FL display notable degradation, characterized by intense heatmap regions and broader error distributions. This suggests the robustness of BD to rotational misalignments commonly encountered in clinical imaging workflows.
In Figure~\ref{fig:Crop_unseen} (crop), where anatomical context is partially removed, BD continues to exhibit strong spatial consistency with low error. Conversely, FL, CC, and TA show larger error regions, particularly along structural boundaries, reflecting their reduced resilience to missing contextual information. Likewise, in Figure S7 (intensity), BD maintains low error across the brain, while TA and other methods exhibit broader and more localized errors. Overall, these findings reinforce BD’s superior robustness to clinically relevant perturbations, making it the most reliable method for deployment in diverse clinical settings. In contrast, TA, FL, and CC show greater sensitivity to such variations, limiting their robustness and practical utility.

 \clearpage
 \subsection{Anatomical Analysis}

\begin{figure}[H]
 \centering
  \includegraphics[width=1.0\textwidth]{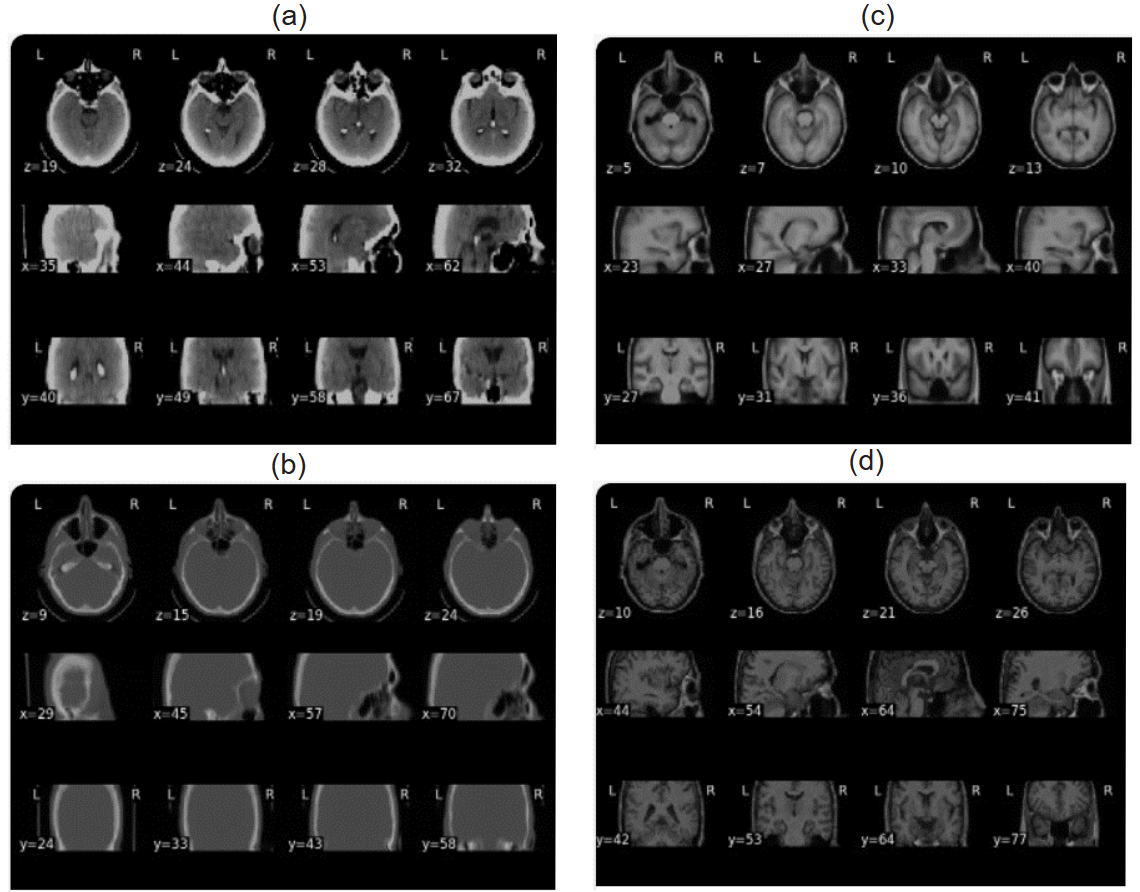}
 \caption{Comparison of (a) windowed CT input highlighting soft tissue structures, (b) original CT input showing the full Hounsfield Unit range, (c) CT-to-T1w translated image generated by the MEAL-BD model, and (d) ground-truth T1w MRI used as reference (using unseen-test data).}
\label{fig:anatomical}
 \end{figure}

 This section qualitatively evaluates the anatomical accuracy of CT-to-T1w translation by comparing the generated images with the input CT scan and ground truth T1w MRI. Figure~\ref{fig:anatomical} illustrates the performance of the proposed MEAL-BD model for CT-to-T1 image translation. Figure~\ref{fig:anatomical}(a) shows the windowed CT input, where soft tissue structures are enhanced by intensity normalization, while panel (b) presents the original CT scan with the full Hounsfield Unit range, which lacks adequate soft tissue contrast. Likewise, Figure~\ref{fig:anatomical}(c) displays the T1w scan generated by the MEAL-BD model, showing improved soft tissue contrast and structural details that closely match the ground truth T1w MRI in Figure~\ref{fig:anatomical}(d). This demonstrates the model’s ability to recover tissue-specific features from CT scans, improving their interpretability in MRI-like contrast space and enabling MRI-informed analysis in settings where MRI is limited or unavailable.

 Anatomically, the generated scans (Figure~\ref{fig:anatomical}c) preserve clinically relevant structures such as the ventricular system, cortical folds, and midline features, which are poorly visualized in the unwindowed CT (Figure~\ref{fig:anatomical}b). The differentiation of gray-white matter in the generated scans closely aligned with the ground-truth MRI (Figure~\ref{fig:anatomical}d), supporting their potential for detecting neurodegenerative changes, tumors, and ischemic lesions. Furthermore, enhanced visibility of the basal ganglia, thalamus, and cortical boundaries suggests that the MEAL-BD model produces diagnostically valuable MRI-like images from CT input. 

\subsection{External Validation}

\begin{figure}[H]
 \centering
  \includegraphics[width=1.0\textwidth]{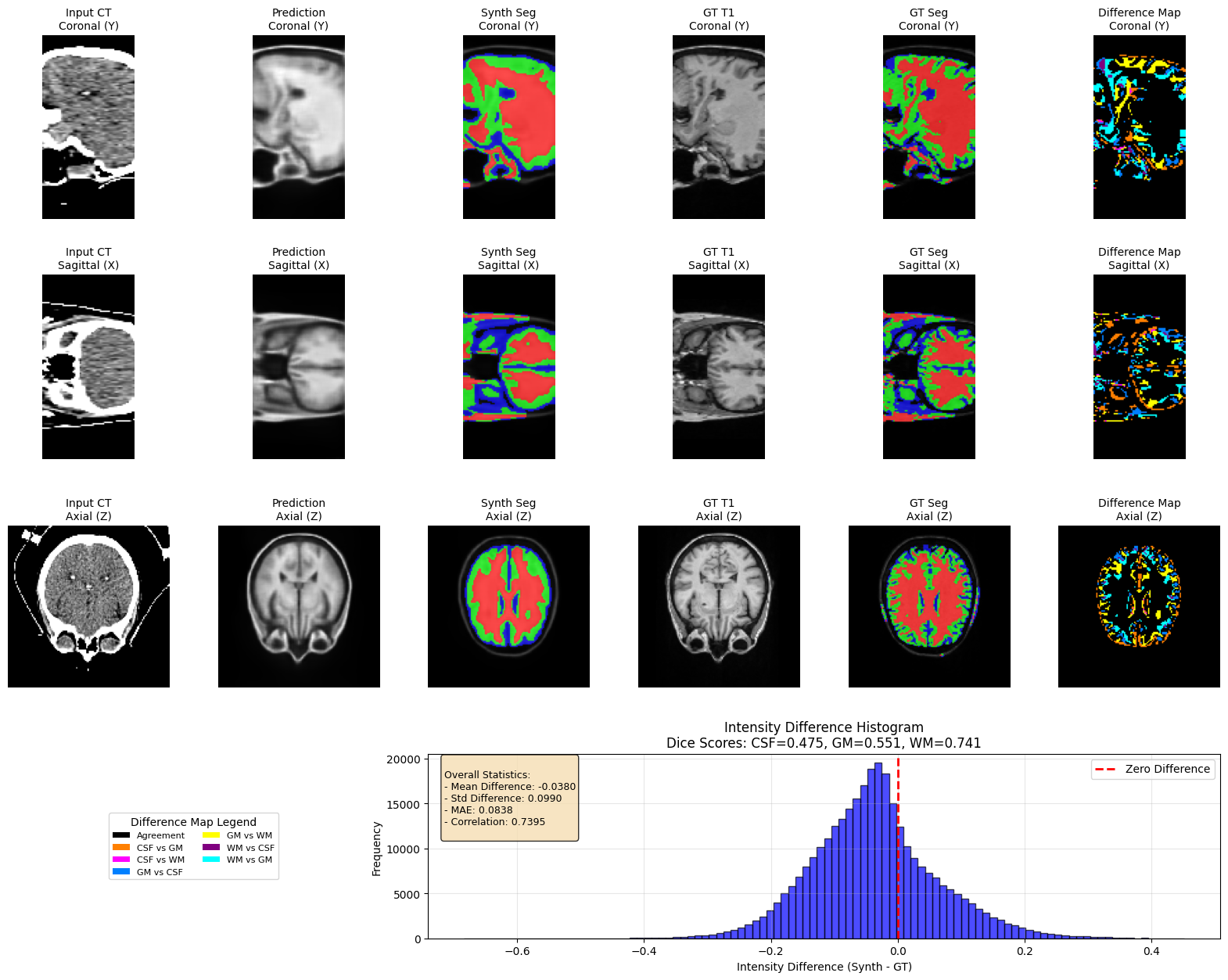}
 \caption{Recovered T1w scan from validation data, segmentations with localized boundary differences and intensity-difference histogram}
\label{fig:Clinical1}
 \end{figure}

Based on 50 validation scans, MEAL demonstrates high anatomical accuracy in synthesizing T1-weighted contrast from CT. The framework achieves a mean PSNR of $\approx$ 23.5 dB and SSIM of $\approx$ 0.80, showing close alignment with ground truth (GT)~\ref{fig:Clinical1}. Tissue-level performance reflects expected contrast availability, with white matter (WM) achieving a Dice of $\approx$ 0.74, gray matter (GM) $\approx$ 0.55, and cerebrospinal fluid (CSF) $\approx$ 0.48 due to limited CT soft-tissue delineation. WM and GM volumes show strong correlation and CSF remains consistent, indicating preserved brain morphology. Qualitative evaluation shows accurate cortical and subcortical reconstruction, and difference maps reveal only localized deviations near tissue boundaries and ventricles. The intensity-difference distribution exhibits a near-zero bias and strong correlation of $r \approx$ 0.74, confirming reliable intensity translation (Figure S20). MEAL therefore yields anatomically coherent synthetic T1w scans with strong WM and GM reconstruction and stable volumetric behavior suitable for downstream neuroimaging tasks.

\subsubsection{Validation with RIRE and PPMI Datasets}

\begin{figure}[H]
 \centering
  \includegraphics[width=1.0\textwidth]{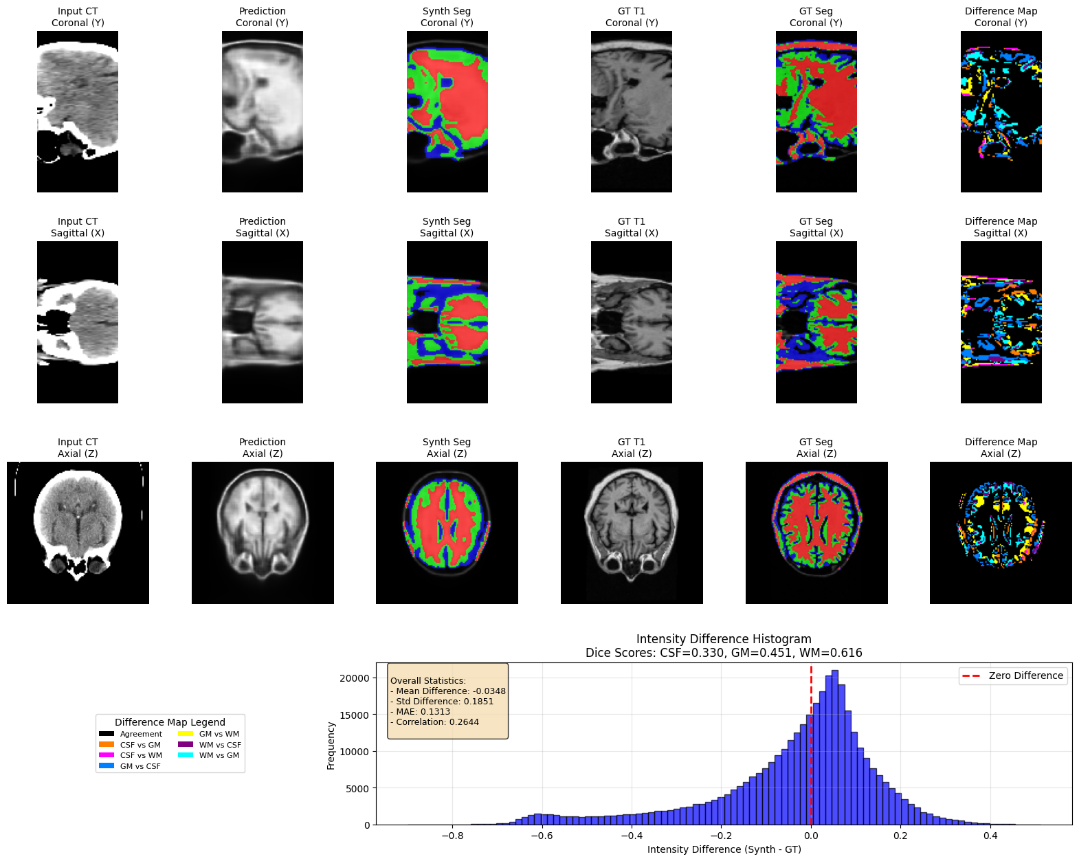}
 \caption{Recovered T1w scan from PPMI data, segmentations with localized boundary differences and intensity-difference histogram}
\label{fig:Clinical2}
 \end{figure}

To assess generalization across diverse clinical settings, we therefore evaluated MEAL on two separate datasets. The RIRE dataset~\cite{west1996comparison, fitzpatrick2002predicting} provides CT–MR pairs suitable for controlled cross-modal evaluation, while the multi-site PPMI dataset introduces real-world variability associated with Parkinson’s disease imaging~\cite{marek2018parkinson,marek2024parkinson}. Using 41 PPMI and 16 RIRE CT scans, MEAL generalizes well to external datasets by recovering T1w scans that preserve overall brain structure but show deviations near ventricles and tissue boundaries. PPMI results show moderate image similarity (PSNR $\approx$ 21.4 dB, SSIM $\approx$ 0.69) and higher tissue segmentation performance (WM $\approx$ 0.62, GM $\approx$ 0.45, CSF $\approx$ 0.33), with a low-positive WM and GM correlation and weaker CSF agreement due to fluid variability, see Figures~\ref{fig:Clinical2} and S21. The RIRE results exhibit lower similarity (PSNR $\approx$ 18.6 dB, SSIM $\approx$ 0.57) and reduced WM segmentation accuracy ($\approx$0.06), while GM ($\approx 0.46$) and CSF ($\approx 0.36$) Dice scores remain comparable to PPMI, see Figures S22 and S23. Both datasets show small intensity bias and reduced correlation, reflecting limited CT soft-tissue contrast, with RIRE also affected by registration\footnote{The difference in contrast between datasets could explain MEAL’s results on RIRE; RIRE CT scans were collected for geometric registration and provide limited soft-tissue detail, whereas PPMI uses standardized biomarker protocols. These acquisition characteristics influence T1 synthesis and contribute to reduced WM separability and image similarity on RIRE.} variability. We must stress that the WM–GM contrast in the RIRE GT T1w scans is insufficiently distinguishable, which limits the performance of conventional segmentation methods and motivates the use of our framework to generate more readily segmentable tissue contrast.

\noindent As previous work shows that MRI-based contrast biomarkers are more sensitive than cortical atrophy in detecting early Alzheimer’s pathology~\cite{putcha2023gray}, our results demonstrate that MEAL can recover clinically relevant T1-like contrast directly from CT across heterogeneous acquisition conditions. MEAL reconstructs T1-like contrast with PSNR $\approx$ 23.5 dB and SSIM $\approx$ 0.80, preserves anatomical patterns on both PPMI and RIRE data, and provides meaningful tissue separability, with WM and GM volumes strongly correlating with MRI GT, enabling the assessment of cortical thickness and GWR (gray-to-white-matter-signal ratio). Since GWR is an MRI-dependent biomarker of early neurodegeneration~\cite{putcha2023gray}, MEAL’s ability to approximate MRI contrast from CT suggests that such biomarkers could be obtained without MRI, broadening access to dementia diagnostics for patients who cannot receive MRI due to cost, contraindications\footnote{Patients may be unable to undergo MRI due to metal implants, cardiac devices, neurostimulators, retained metal fragments, severe claustrophobia~\cite{carr2002magnetic,shellock2000magnetic}, inability to remain still~\cite{hand2005magnetic}, or critical illness requiring equipment incompatible with MRI~\cite{shellock2000magnetic}.}, limited availability, or urgent clinical settings

\section{Qualitative Comparison with Augmentation-Aware Models}

Table S2 presents a comparison of the MEAL-BD framework with several augmentation-aware learning methods, including CASSLE, AugSelf, LooC~\cite{xiao2020should}, and IFM~\cite{robinson2021can}. The analysis covers application domains, architectural designs, augmentation strategies, training objectives, and key strengths. MEAL-BD excels in the medical imaging domain, particularly for CT-to-MRI translation, employing a multi-encoder architecture with a dynamic controller per augmentation type. It uses supervised learning with adaptive fusion and demonstrates strong generalization, robustness to scanner and patient variability, and effective anatomical feature learning. In contrast, methods like CASSLE and AugSelf target natural images and follow self-supervised learning paradigms. CASSLE employs a conditioned projector with minimal architectural change, while AugSelf adds a network to predict augmentation parameters. LooC introduces a more complex multi-projector and multi-loss scheme for disentangled learning, albeit with increased computational cost. IFM adopts a simpler masking-based SSL approach without explicit augmentation input. Overall, MEAL-BD is well-suited for medical applications, balancing protocol invariance and structural fidelity, while the other methods focus on general-purpose augmentation-aware representation learning. Although the tasks and input domains differ, the comparison reveals transferable design principles for managing augmentation-induced variability. Quantitative metrics are not compared, as each method is tailored to distinct goals and data types.

\begin{center}
\small
\begin{longtable}{>{\raggedright\arraybackslash}p{3.3cm} >{\raggedright\arraybackslash}p{6.5cm} >{\raggedright\arraybackslash}p{6.5cm}}
\caption{Qualitative comparison of MEAL and BBDM-SKC+ISTA~\cite{choo2024slice} Frameworks} \label{tab:meal_vs_bbdm} \\
\toprule
\textbf{Frameworks} & \textbf{MEAL} & \textbf{BBDM-SKC + ISTA} \\
\midrule
\textbf{Goal} & Robust CT-to-MRI translation under real-world clinical variability (augmentations) & Consistent and deterministic 3D MRI synthesis from 2D slices \\
\addlinespace
\textbf{Architecture} & Multi-encoder U-Net with 4 augmentation-specific branches and a dynamic controller block & Single 2D Brownian Bridge (DM) with no extra models but enhanced with  Style Key Conditioning (SKC) and Inter-Slice Trajectory Alignment (ISTA) for 3D consistency \\
\addlinespace
\textbf{Fusion Strategy} & Attention-based fusion using learned weights ($\alpha_k$) from a controller network (adaptive to augmentations) & No fusion layer \\
\addlinespace
\textbf{Augmentation View} & Treats augmentation as learnable modules in the graph (rotation, flip, crop, intensity); preserves augmentation-specific features & No explicit augmentation or data diversity explored; merely assumes paired CT-MRI consistency \\
\addlinespace
\textbf{3D Consistency} & Ensured via end-to-end learned representation over full 3D volume and controller-guided feature fusion & Ensured via ISTA which corrects direction between slices during denoising \\
\addlinespace
\textbf{Style Consistency} & Achieved by consistent input distribution control and training with augmentations & Achieved via SKC using target MRI histograms \\
\addlinespace
\textbf{Determinism} & End-to-end deterministic (except for augmentation randomness in training); controller helps regularize & Fully deterministic generation using Brownian Bridge formulation \\
\addlinespace
\textbf{Structural Preservation} & Superior SSIM under perturbations; anatomical structures preserved due to dynamic weighting of relevant views & High structural integrity preserved across slices due to ISTA + BBDM’s stable noise modeling \\
\addlinespace
\textbf{Adaptability to Clinical Variability} & MEAL-BD strongly adapts to rotation, flip, crop, intensity variance (verified statistically) & not tested with augmentations; more focused on internal consistency than external variability \\
\addlinespace
\textbf{Model Size} & Large ($\sim$273M parameters); heavy due to parallel augmentation streams & Lightweight (single 2D DM); no additional encoder/decoder models used \\
\addlinespace
\textbf{Generalization} & Generalizes well to unseen augmented data; superior in both PSNR/SSIM across all test types & Generalizes well to different 3D brain structures; relies on style conditioning and inter-slice modeling \\
\addlinespace
\textbf{Statistical Validation} & Strong: Kruskal--Wallis and Dunn’s post-hoc tests confirm BD $>$ FL, CC, TA & Strong: Quantitative (PSNR, SSIM, NRMSE) + qualitative results outperform 2D and 3D baselines \\
\addlinespace
\textbf{Modularity} & Highly modular (encoders can be pruned/adapted); pip-installable & Minimal modularity, focused on elegant improvement to diffusion process \\
\bottomrule
\end{longtable}
\end{center}

For medical imaging, MEAL-BD offers a robust framework for CT-to-MRI translation that effectively handles real-world clinical variability. However, BBDM-SKC+ISTA performs CT-to-MRI translation without explicitly accounting for this variability (see Table~\ref{tab:meal_vs_bbdm}). MEAL employs a multi-stream encoder architecture with augmentation-specific branches and a dynamic attention-based fusion mechanism, allowing it to adapt to the challenges pose by various perturbation. BBDM-SKC+ISTA\footnotetext{
Quantitative comparison with BBDM-SKC+ISTA is precluded by differences in model formulation and the lack of publicly available pretrained weights/model and training code. 
} ,on the other hand, uses a single, lightweight 2D Brownian Bridge Diffusion Model enhanced with SKC and ISTA, focusing on slice-wise consistency and structural coherence across 3D volumes. While MEAL excels in adaptability and generalization under augmentation-induced perturbations, BBDM offers deterministic, style-consistent synthesis with minimal architectural overhead.

\subsection{Discussion}

Our results demonstrate the robustness of data-driven methods in medical image reconstruction. Without perturbation, BD achieved the highest performance (PSNR = 23.03; SSIM = 0.73 in unseen data), outperforming CC (PSNR = 22.41; SSIM = 0.70) despite the integration of multistream features. FL also performed well through spatial-frequency fusion, but slightly trailed BD. TA consistently showed the poorest performance, with significant spatial degradation. These results highlight the advantage of robust, learned-representations over simpler, task-agnostic strategies. Notably, BD maintained superior generalization under clinically relevant augmentations consistently reducing reconstruction errors. This establishes BD as the most reliable method across both unseen- and predefined-test sets for deployment in variable clinical settings.

\subsubsection{Metric Sensitivity and Clinical Relevance}

A key insight from our study is the complementary value of PSNR and SSIM in assessing medical image quality. Although both metrics showed significant differences across methods under geometric and intensity augmentations (p-value $< \alpha$ (0.01), see Table~\ref{table:unseen_data_summary}), SSIM was more sensitive to structural fidelity. For instance, TA achieved moderate PSNR but low SSIM, indicating poor anatomical preservation despite pixel-level similarity. In contrast, BD consistently achieved the highest SSIM, demonstrating superior structural integrity. These findings underscore the importance of including perceptually aligned metrics like SSIM in medical imaging, where structural preservation is essential for clinical reliability.

\subsubsection{Toward Hybrid Models and Future Directions}

Despite their strengths, CC and FL remained sensitive to geometric transformations such as flipping and rotation (Figures~\ref{fig:Rotation_unseen} and \ref{fig:Crop_unseen}), with error heatmaps revealing localized distortions, particularly around cortical structures. In contrast, BD consistently maintained low reconstruction errors across all tested perturbations, showing no notable degradation in spatial fidelity. This robustness positions BD as the most reliable choice, though future work could explore hybrid designs that combine CC’s multistream encoding and FL’s spatial-frequency fusion with BD’s geometric resilience. Additionally, integrating domain-specific augmentations, such as simulated lesions or scanner-induced variations, may further strengthen generalizability without compromising anatomical accuracy.

\subsubsection{Performance Gains of BD in Multi-Stream Augmentation}

The proposed multi-stream architecture explores distinct strategies for fusing augmentation-specific features, with the BD variant introducing key innovations to preserve augmentation-aware representations. While CC and FL process pre-augmented inputs through separate encoders followed by static fusion using concatenation or 1×1 convolutions, BD uniquely embeds augmentation operations as differentiable layers within a unified computational graph. This design enables end-to-end learning of augmentation-specific feature interactions, enhancing representational expressiveness and adaptability.

Given the input volume $\mathbf{X} \in \mathbb{R}^{H \times W \times D \times C}$ propagates through four parallel augmentation streams $\{A_k(\mathbf{X})\}_{k=1}^4$, each processed by a shared encoder \(f_{\theta}\). The BD controller network then computes dynamic attention weights using ~\ref{eqn:alpha}. The fused representation in equation~\ref{eqn:BD} explicitly preserves augmentation-specific patterns through differentiable weighting, unlike CC and FL, which rely on fixed fusion mechanisms. This is followed by a parameter-efficient decoder that reconstructs spatial features using $\hat{\mathbf{F}} = \Gamma_{\phi}(\boldsymbol{\mathbf{F_{\rm BD}}})$ so that $\Gamma_{\phi}$ employs residual transpose convolutions iteratively refined as defined in ~\ref{eqn:ResidualTranspose}. By jointly optimizing the parameter $(\theta, \phi, \mathbf{W})$ using gradient descent, BD enables augmentation-equivalent feature learning by adaptively reweighting streams based on their semantic relevance, whereas CC and FL treat augmentations as static, non-adaptive inputs.

This differentiable architecture not only reduces parameter overhead (due to the shared encoder $f_{\theta}$, in contrary to CC and FL's independent encoders), but also supports interpretable analysis of augmentation contributions through $\alpha_k$, a critical advantage for domain-specific applications that require explicit attribution of learned transformations.

\subsubsection{Clinical Relevance and Broader Implications}

MEAL addresses a critical challenge in medical AI, balancing data augmentation with robust feature learning. By treating augmentations as alternative anatomical views, MEAL builds protocol-agnostic representations, moving beyond conventional brute-force augmentation strategies. Its controller block dynamically weighs features (Figure~\ref{fig:ML}), analogous to a radiologist’s emphasis on clinically relevant cues such as edge definition for tumor margins or intensity consistency for tissue contrast. This design supports improved structural preservation and positions MEAL as a practical tool for enhancing medical image analysis workflows. In particular, MEAL (BD) demonstrates potential value in resource-limited settings where MRI access is constrained but soft-tissue contrast remains clinically important.

Beyond visual image translation, MEAL enables clinically relevant tissue-level analysis. The synthesized T1w MRI-like scans support white WM and GM segmentation and volumetric quantification, which are routinely used in the clinical assessment of neurodegenerative diseases~\cite{foo2016progression,habes2018regional}, brain atrophy~\cite{dadar2021white}, and structural abnormalities~\cite{debette2010clinical}. In our experiments, WM and GM segmentation derived from MEAL-generated images showed strong agreement with ground-truth MRI, indicating preserved tissue contrast and anatomical consistency. This capability suggests that MEAL could facilitate quantitative neuroimaging analysis in clinical scenarios where MRI is not available, extending the utility of CT beyond conventional anatomical visualization. 

A major concern in medical image translation is hallucination, in which synthesized images introduce anatomically plausible but clinically incorrect structures, an issue that has justified the limited clinical adoption of translation-based methods~\cite{kim2024tackling}. Consequently, MEAL is explicitly designed to mitigate hallucination rather than to maximize perceptual realism. First, MEAL is trained exclusively on paired CT–MRI data, which constrains the model to learn anatomically grounded mappings instead of unconstrained generation. Second, the framework prioritizes structural similarity (SSIM), tissue segmentation consistency, and voxel-wise error analysis over perceptual sharpness, thus reducing incentives for hallucinated detail. Third, the augmentation-aware controller adaptively suppresses transformations that compromise anatomical fidelity, as reflected by consistently lower error maps under geometric and intensity perturbations.

Importantly, the synthesized MRI-like scans are intended to support clinical interpretation and downstream analysis (such as segmentation, morphometry, and contrast approximation), serving as an adjunct to the radiologist's assessment, rather than as a direct replacement for acquired MRI or as a standalone diagnostic tool. Regions associated with higher uncertainty, such as tissue boundaries and ventricular areas, are explicitly reflected in error maps, providing transparency and aiding informed clinical interpretation.

\section{Remarks and Future Directions}

Despite the demonstrated effectiveness of the proposed framework, there remain avenues for further improvement. Enhanced inter-branch connectivity is needed to support efficient skip connections in the decoder, as poorly coordinated pathways may hinder gradient flow and multi-scale feature integration. Additionally, each new encoder increases parameter count linearly, posing memory challenges, especially in high-resolution 3D imaging. These considerations highlights the need for empirical studies to balance model complexity with computational efficiency.

Future investigations should aim to identify the optimal number of encoder heads that maximizes performance while minimizing computational overhead. Performance saturation, where additional heads yield diminishing returns, alongside memory-throughput trade-offs, should be systematically assessed in relation to GPU constraints and batch size. To further promote the specialization of features within each augmentation stream, head-specific loss functions tailored to transformation types (e.g., contrastive loss for intensity augmentations, Dice loss for spatial transformations) may be beneficial. In addition, gradient balancing strategies could be explored to ensure stable and effective learning across heterogeneous loss functions. Another promising direction involves the development of a dynamic gating mechanism that adaptively prunes or combines encoder heads during training, enabling more efficient resource utilization without compromising representational capacity.

This study focuses on CT-to-T1w MRI as a case study to demonstrate the unique capabilities of the MEAL pipeline. Generalization to other pair of modality will require additional validation and task-specific tuning. Furthermore, synthesized MRI-like scans are not intended to replace acquired MRI for diagnostic purposes, but rather to serve as a decision-support tool that complements radiologist assessment, particularly in resource-limited settings. Finally, the multi-encoder architecture introduces increased computational cost during training; however, inference involves only the trained model and is substantially less demanding, though further optimization may still be required for large-scale deployment.

Future work should explore hybrid architectures that balance fidelity, robustness, and generalizability by integrating CC’s multistream features, FL’s spatial-frequency fusion, and BD’s geometric invariance. Incorporating domain-specific augmentations and extending MEAL to multi-modal fusion (e.g., PET-MRI) and federated learning could further enhance performance and clinical applicability.

\section{Conclusions}

This work introduced MEAL, a multi-encoder augmentation-aware learning framework for robust CT-to-T1w MRI translation. Across extensive quantitative, qualitative, and statistical evaluations, the BD variant consistently outperformed competing single-stream and multi-stream baselines, achieving superior PSNR and SSIM while maintaining exceptional robustness under geometric and intensity perturbations. By reframing data augmentation as a source of complementary anatomical views rather than stochastic noise, MEAL enables the learning of protocol-invariant representations that remain stable under clinically realistic variability. The dynamic controller-driven fusion mechanism allows the model to selectively emphasize informative transformations while suppressing detrimental ones, resulting in improved structural fidelity and reliable anatomical reconstruction.

Importantly, MEAL demonstrates strong generalization to external datasets and preserves clinically relevant tissue contrast, supporting downstream tasks such as segmentation and morphometric analysis. This capability highlights MEAL’s potential to extend MRI-like diagnostic insights to CT-only settings, particularly in resource-limited or time-critical clinical environments. With its modular, interpretable design and compatibility with standard medical imaging formats, MEAL provides a principled foundation for augmentation-aware medical AI systems. Beyond CT-to-MRI translation, the framework is readily extensible to other imaging modalities and tasks, positioning MEAL as a general solution for robust and generalizable medical image learning under real-world variability.

% This work presents MEAL, a novel and generalizable framework for CT-to-T1w MRI translation. Across all evaluations, BD consistently achieved superior performance, balancing high PSNR and SSIM with exceptional robustness to geometric and intensity perturbations. These results highlight MEAL’s clinical potential for tasks such as medical intervention and surgical planning, particularly in MRI-limited settings.

% By rethinking data augmentation as a mechanism for learning protocol-invariant representations, MEAL improves translation reliability under real-world variability. Its integration of multistream encoders, dynamic feature fusion, and diagnostic evaluation enables structural fidelity while maintaining compatibility with standard medical formats (NIfTI/DICOM). With its modular and interpretable design, MEAL provides a strong foundation for robust, generalizable AI solutions across diverse clinical environments.

\section*{Acknowledgements}

The authors would like to acknowledge the financial support from the Research Grants Council of Hong Kong (Grants: 11102218, 11200422, RFS2223-1S02, C1134-20G), City University of Hong Kong (Grants: 7005626, 7030012, 9609321, and 9610616), the Tung Biomedical Sciences Centre, the State Key Laboratory of Terahertz and Millimeter Waves, the Innovation and Technology Commission – InnoHK – Hong Kong Centre for Cerebro-cardiovascular Health Engineering, the Midstream Research Programme for Universities (MHKJS: MHP/076/23), and the National Key Research and Development Program of China (Grant: 2023YFE0210300).

\section*{Declaration of Interest}
The authors declare that there is no any competing interests.

\section*{\it Code availability}

\begin{enumerate}
    \item Relevant models are available on \textcolor{blue}{\href{https://huggingface.co/AI-vBRAIN/pyMEAL}{Hugging Face}}.
    
    \item The source code, packaged as \texttt{pip}-installable software, is available on \textcolor{blue}{\href{https://pypi.org/project/pyMEAL/}{PyPI}}.
    
    \item Relevant code and tutorials are available on \textcolor{blue}{\href{https://github.com/ai-vbrain/pyMEAL}{GitHub}}.
    \item Relevant \textcolor{blue}{\href{https://github.com/ai-vbrain/pyMEAL/tree/main/docs}{supplementary material}} is available on \textcolor{blue}{\href{https://github.com/ai-vbrain/pyMEAL}{GitHub}}
\end{enumerate}

% Relevant code is also available at

\section*{Credit Authorship Contribution}

Abdul-mojeed Olabisi Ilyas: Software, Conceptualization, Methodology, Data Curation, Formal Analysis, Validation, Investigation, Visualization, Writing – review $\&$ editing.  Adeleke Maradesa: Software, Methodology, Data Curation, Formal Analysis, Investigation, deployment, Visualization, Writing – original draft, review and editing. Jamal Banzi: Investigation, Writing – review $\&$ editing. Jianpan  Huang: Investigation, Writing – review $\&$ editing. Henry K.F. Mak: Formal Analysis, Writing – review $\&$ editing. Kannie W.Y: Writing – review $\&$ editing, Supervision, Resources, Project administration, Investigation, Funding acquisition.

\section*{Declaration of AI-assisted technologies in the revision process}

To improve grammatical coherence, clarity and readability, Groq and Gemini Advanced were used to assist in the revision process. The authors subsequently conducted a thorough review, refinement, and finalization of the manuscript, assuming full responsibility for its content.

\clearpage
\section*{List of Symbols and Abbreviations}
 \FloatBarrier
\begin{table}[h!]
\centering
\renewcommand{\arraystretch}{1.3}
\begin{tabularx}{1.0\textwidth}{|c|X|}
\hline
\textbf{Symbol} & \textbf{Definition} \\
\hline
$\mathbf{X}$ & Input image volume or tensor \\
$f_{\theta_k}, f_{\theta_k'}$ & Encoder with learnable parameters $\theta_k$ (per stream) or $\theta_k'$ (non-shared) \\
$R_f(\cdot)$ & Refined residual block with filter size $f$ \\
$\mathbf{F}_{\text{FL}}$ & Fused feature map from Fusion Layer (FL) \\
$\mathbf{F}_{\text{cc}}$ & Fused feature map from Concatenative Fusion (CC) \\
$\mathbf{F}_{\text{BD}}$ & Fused feature map from Build Controller (BD) model \\
$A_k$ & Differentiable augmentation module (e.g., flip, rotate, crop, intensity shift) for stream $k$ \\
$\alpha_k$ & Attention weight for the $k^{th}$ augmented stream, dynamically learned \\
$\mathbf{w}, \mathbf{W}$ & Trainable projection weights for attention computation \\
$\mathrm{GAP}(\cdot)$ & Global average pooling operator \\
$\Gamma_\phi$ & Decoder network with parameters $\phi$ \\
$\mathcal{U}$ & Upsampling layer (e.g., nearest neighbor or transposed convolution) \\
$h_k$ & Encoded feature map from the $k^{th}$ stream \\
$\bigoplus$ & Channel-wise concatenation operator \\
$C_{\text{fuse}}$ & 1×1×1 convolution bottleneck for fusion \\
$\hat{\mathbf{F}}$ & Final decoded feature representation \\
$\mathrm{BD}$ & Builder Block \\
$\mathrm{CC}$ & Encoder Concatenation \\
$\mathrm{FL}$ & Fusion Layer \\
$\mathrm{TA}$ & Traditional Augmentation \\
$\mathrm{NA}$ & No Augmentation \\
$\mathrm{MRI}$ & Magnetic Resonance Imaging \\
$\mathrm{CT}$ & Computed Tomography \\
$\mathrm{MEAL}$ & Multi-Encoder Augmentation-Aware
Learning \\
$\mathrm{pyMEAL}$ & Python-based implementation of MEAL \\
$\mathrm{PSNR}$ & Peak Signal-to-Noise Ratio \\
$\mathrm{SSIM}$ & Structural Similarity Index Measure \\
$\mathrm{DL}$ & Deep Learning \\
$\mathrm{RRBs}$ & Refined Residual Blocks \\
$\mathrm{GM}$ & Grey matter \\
$\mathrm{WM}$ & White Matter \\
\hline
\end{tabularx}
\label{tab:symbols_abbrev}
\end{table}

\clearpage
%\bibliography{REF}

\begin{thebibliography}{10}
\expandafter\ifx\csname url\endcsname\relax
  \def\url#1{\texttt{#1}}\fi
\expandafter\ifx\csname urlprefix\endcsname\relax\def\urlprefix{URL }\fi
\expandafter\ifx\csname href\endcsname\relax
  \def\href#1#2{#2} \def\path#1{#1}\fi

\bibitem{hussain2022modern}
S.~Hussain, I.~Mubeen, N.~Ullah, S.~S. U.~D. Shah, B.~A. Khan, M.~Zahoor,
  R.~Ullah, F.~A. Khan, M.~A. Sultan, Modern diagnostic imaging technique
  applications and risk factors in the medical field: a review, BioMed research
  international 2022~(1) (2022) 5164970.

\bibitem{fountzilas2025convergence}
E.~Fountzilas, T.~Pearce, M.~A. Baysal, A.~Chakraborty, A.~M. Tsimberidou,
  Convergence of evolving artificial intelligence and machine learning
  techniques in precision oncology, npj Digital Medicine 8~(1) (2025) 75.

\bibitem{wagner1991advances}
H.~N. Wagner~Jr, P.~S. Conti, Advances in medical imaging for cancer diagnosis
  and treatment, Cancer 67~(S4) (1991) 1121--1128.

\bibitem{attariwala2013whole}
R.~Attariwala, W.~Picker, Whole body mri: improved lesion detection and
  characterization with diffusion weighted techniques, Journal of Magnetic
  Resonance Imaging 38~(2) (2013) 253--268.

\bibitem{cossio2023augmenting}
M.~Cossio, Augmenting medical imaging: a comprehensive catalogue of 65
  techniques for enhanced data analysis, arXiv preprint arXiv:2303.01178
  (2023).

\bibitem{ostertagova2014methodology}
E.~Ostertagova, O.~Ostertag, J.~Kov{\'a}{\v{c}}, Methodology and application of
  the kruskal-wallis test, Applied mechanics and materials 611 (2014) 115--120.

\bibitem{alshardan2024leveraging}
A.~Alshardan, N.~Alruwais, H.~Alqahtani, A.~Alshuhail, W.~S. Almukadi,
  A.~Sayed, Leveraging transfer learning-driven convolutional neural
  network-based semantic segmentation model for medical image analysis using
  mri images, Scientific Reports 14~(1) (2024) 30549.

\bibitem{chaari2025hybrid}
A.~Chaari, I.~Fourati~Kallel, S.~Kammoun, M.~Frikha, Hybrid data augmentation
  strategies for robust deep learning classification of corneal topographic
  maptopographic map, Biomedical Physics \& Engineering Express (2025).

\bibitem{alzubaidi2021review}
L.~Alzubaidi, J.~Zhang, A.~J. Humaidi, A.~Al-Dujaili, Y.~Duan, O.~Al-Shamma,
  J.~Santamar{\'\i}a, M.~A. Fadhel, M.~Al-Amidie, L.~Farhan, Review of deep
  learning: concepts, cnn architectures, challenges, applications, future
  directions, Journal of big Data 8 (2021) 1--74.

\bibitem{elton2019deep}
D.~C. Elton, Z.~Boukouvalas, M.~D. Fuge, P.~W. Chung, Deep learning for
  molecular design—a review of the state of the art, Molecular Systems Design
  \& Engineering 4~(4) (2019) 828--849.

\bibitem{huang2023self}
S.-C. Huang, A.~Pareek, M.~Jensen, M.~P. Lungren, S.~Yeung, A.~S. Chaudhari,
  Self-supervised learning for medical image classification: a systematic
  review and implementation guidelines, NPJ Digital Medicine 6~(1) (2023) 74.

\bibitem{mumuni2022data}
A.~Mumuni, F.~Mumuni, Data augmentation: A comprehensive survey of modern
  approaches, Array 16 (2022) 100258.

\bibitem{alomar2023data}
K.~Alomar, H.~I. Aysel, X.~Cai, Data augmentation in classification and
  segmentation: A survey and new strategies, Journal of Imaging 9~(2) (2023)
  46.

\bibitem{yang2023survey}
Z.~Yang, R.~O. Sinnott, J.~Bailey, Q.~Ke, A survey of automated data
  augmentation algorithms for deep learning-based image classification tasks,
  Knowledge and Information Systems 65~(7) (2023) 2805--2861.

\bibitem{garcea2023data}
F.~Garcea, A.~Serra, F.~Lamberti, L.~Morra, Data augmentation for medical
  imaging: A systematic literature review, Computers in Biology and Medicine
  152 (2023) 106391.

\bibitem{bosquet2023full}
B.~Bosquet, D.~Cores, L.~Seidenari, V.~M. Brea, M.~Mucientes, A.~Del~Bimbo, A
  full data augmentation pipeline for small object detection based on
  generative adversarial networks, Pattern Recognition 133 (2023) 108998.

\bibitem{kaji2019overview}
S.~Kaji, S.~Kida, Overview of image-to-image translation by use of deep neural
  networks: denoising, super-resolution, modality conversion, and
  reconstruction in medical imaging, Radiological physics and technology 12~(3)
  (2019) 235--248.

\bibitem{shorten2019survey}
C.~Shorten, T.~M. Khoshgoftaar, A survey on image data augmentation for deep
  learning, Journal of big data 6~(1) (2019) 1--48.

\bibitem{kebaili2023deep}
A.~Kebaili, J.~Lapuyade-Lahorgue, S.~Ruan, Deep learning approaches for data
  augmentation in medical imaging: a review, Journal of imaging 9~(4) (2023)
  81.

\bibitem{chlap2021review}
P.~Chlap, H.~Min, N.~Vandenberg, J.~Dowling, L.~Holloway, A.~Haworth, A review
  of medical image data augmentation techniques for deep learning applications,
  Journal of medical imaging and radiation oncology 65~(5) (2021) 545--563.

\bibitem{abdollahi2020data}
B.~Abdollahi, N.~Tomita, S.~Hassanpour, Data augmentation in training deep
  learning models for medical image analysis, Deep learners and deep learner
  descriptors for medical applications (2020) 167--180.

\bibitem{goceri2023medical}
E.~Goceri, Medical image data augmentation: techniques, comparisons and
  interpretations, Artificial Intelligence Review 56~(11) (2023) 12561--12605.

\bibitem{przewikezlikowski2024augmentation}
M.~Przewi{\k{e}}{\'z}likowski, M.~Pyla, B.~Zieli{\'n}ski, B.~Twardowski,
  J.~Tabor, M.~{\'S}mieja, Augmentation-aware self-supervised learning with
  conditioned projector, Knowledge-Based Systems 305 (2024) 112572.

\bibitem{kim2025augward}
M.~Kim, J.~Choi, S.~Lee, J.~Jung, U.~Kang, Augward: Augmentation-aware
  representation learning for accurate graph classification, arXiv preprint
  arXiv:2503.21105 (2025).

\bibitem{trzcinski2024zero}
T.~Trzcinski, B.~Twardowski, B.~Zieli{\'n}ski, K.~Adamczewski, B.~W{\'o}jcik,
  Zero-waste machine learning, in: ECAI 2024, IOS Press, 2024, pp. 43--49.

\bibitem{sandfort2019data}
V.~Sandfort, K.~Yan, P.~J. Pickhardt, R.~M. Summers, Data augmentation using
  generative adversarial networks (cyclegan) to improve generalizability in ct
  segmentation tasks, Scientific reports 9~(1) (2019) 16884.

\bibitem{rasool2025pixmed}
M.~A. Rasool, A.~Abdusalomov, A.~Kutlimuratov, M.~A. Ahamed, S.~Mirzakhalilov,
  A.~Shavkatovich~Buriboev, H.~S. Jeon, Pixmed-enhancer: An efficient approach
  for medical image augmentation, Bioengineering 12~(3) (2025) 235.

\bibitem{yang2020mri}
Q.~Yang, N.~Li, Z.~Zhao, X.~Fan, E.~I.-C. Chang, Y.~Xu, Mri cross-modality
  image-to-image translation, Scientific reports 10~(1) (2020) 3753.

\bibitem{choo2024slice}
K.~Choo, Y.~Jun, M.~Yun, S.~J. Hwang, Slice-consistent 3d volumetric brain
  ct-to-mri translation with 2d brownian bridge diffusion model, in:
  International Conference on Medical Image Computing and Computer-Assisted
  Intervention, Springer, 2024, pp. 657--667.

\bibitem{chen2019differentiation}
H.~Chen, Y.~Zhang, J.~Pang, Z.~Wu, M.~Jia, Q.~Dong, W.~Xu, The differentiation
  of soft tissue infiltration and surrounding edema in an animal model of
  malignant bone tumor: evaluation by dual-energy ct, Technology in cancer
  research \& treatment 18 (2019) 1533033819846842.

\bibitem{bearcroft2007imaging}
P.~W. Bearcroft, Imaging modalities in the evaluation of soft tissue
  complaints, Best Practice \& Research Clinical Rheumatology 21~(2) (2007)
  245--259.

\bibitem{lamontagne2019oasis}
P.~J. LaMontagne, T.~L. Benzinger, J.~C. Morris, S.~Keefe, R.~Hornbeck,
  C.~Xiong, E.~Grant, J.~Hassenstab, K.~Moulder, A.~G. Vlassenko, et~al.,
  Oasis-3: longitudinal neuroimaging, clinical, and cognitive dataset for
  normal aging and alzheimer disease, medrxiv (2019) 2019--12.

\bibitem{shinohara2014statistical}
R.~T. Shinohara, E.~M. Sweeney, J.~Goldsmith, N.~Shiee, F.~J. Mateen, P.~A.
  Calabresi, S.~Jarso, D.~L. Pham, D.~S. Reich, C.~M. Crainiceanu, et~al.,
  Statistical normalization techniques for magnetic resonance imaging,
  NeuroImage: Clinical 6 (2014) 9--19.

\bibitem{case2008fundamentals}
J.~A. Case, B.~L. Hsu, S.~J. Cullom, Fundamentals of computed tomography and
  computed tomography angiography, Nuclear Cardiology: Technical Applications
  (2008) 249.

\bibitem{abadi2016tensorflow}
M.~Abadi, P.~Barham, J.~Chen, Z.~Chen, A.~Davis, J.~Dean, M.~Devin,
  S.~Ghemawat, G.~Irving, M.~Isard, et~al., $\{$TensorFlow$\}$: a system for
  $\{$Large-Scale$\}$ machine learning, in: 12th USENIX symposium on operating
  systems design and implementation (OSDI 16), 2016, pp. 265--283.

\bibitem{tustison2024antsx}
N.~J. Tustison, M.~A. Yassa, B.~Rizvi, P.~A. Cook, A.~J. Holbrook, M.~T.
  Sathishkumar, M.~G. Tustison, J.~C. Gee, J.~R. Stone, B.~B. Avants, Antsx
  neuroimaging-derived structural phenotypes of uk biobank, Scientific Reports
  14~(1) (2024) 8848.

\bibitem{tustison2021antsx}
N.~J. Tustison, P.~A. Cook, A.~J. Holbrook, H.~J. Johnson, J.~Muschelli, G.~A.
  Devenyi, J.~T. Duda, S.~R. Das, N.~C. Cullen, D.~L. Gillen, et~al., The antsx
  ecosystem for quantitative biological and medical imaging, Scientific reports
  11~(1) (2021) 9068.

\bibitem{montalt2022tensorflow}
J.~Montalt-Tordera, J.~Steeden, V.~Muthurangu, Tensorflow mri: a library for
  modern computational mri on heterogenous systems, in: Proceedings of the 31st
  Annual Meeting of ISMRM, London, UK, 2022, p. 2769.

\bibitem{raina2023tackling}
V.~Raina, N.~Molchanova, M.~Graziani, A.~Malinin, H.~Muller, M.~B. Cuadra,
  M.~Gales, Tackling bias in the dice similarity coefficient: introducing ndsc
  for white matter lesion segmentation, in: 2023 IEEE 20th International
  Symposium on Biomedical Imaging (ISBI), IEEE, 2023, pp. 1--5.

\bibitem{razali2011power}
N.~M. Razali, Y.~B. Wah, et~al., Power comparisons of shapiro-wilk,
  kolmogorov-smirnov, lilliefors and anderson-darling tests, Journal of
  statistical modeling and analytics 2~(1) (2011) 21--33.

\bibitem{rosner2006wilcoxon}
B.~Rosner, R.~J. Glynn, M.-L.~T. Lee, The wilcoxon signed rank test for paired
  comparisons of clustered data, Biometrics 62~(1) (2006) 185--192.

\bibitem{ruxton2008time}
G.~D. Ruxton, G.~Beauchamp, Time for some a priori thinking about post hoc
  testing, Behavioral ecology 19~(3) (2008) 690--693.

\bibitem{seabold2010statsmodels}
S.~Seabold, J.~Perktold, Statsmodels: econometric and statistical modeling with
  python., SciPy 7~(1) (2010) 92--96.

\bibitem{virtanen2020scipy}
P.~Virtanen, R.~Gommers, T.~E. Oliphant, M.~Haberland, T.~Reddy, D.~Cournapeau,
  E.~Burovski, P.~Peterson, W.~Weckesser, J.~Bright, et~al., Scipy 1.0:
  fundamental algorithms for scientific computing in python, Nature methods
  17~(3) (2020) 261--272.

\bibitem{terpilowski2019scikit}
M.~A. Terpilowski, scikit-posthocs: Pairwise multiple comparison tests in
  python, Journal of Open Source Software 4~(36) (2019) 1169.

\bibitem{laha2018skull}
M.~Laha, P.~C. Tripathi, S.~Bag, A skull stripping from brain mri using
  adaptive iterative thresholding and mathematical morphology, in: 2018 4th
  International Conference on Recent Advances in Information Technology (RAIT),
  IEEE, 2018, pp. 1--6.

\bibitem{sang2024improved}
G.~Sang, X.~Wang, J.~Zhang, Improved otsu theory of image multi-threshold
  segmentation by incorporating ant colony algorithm, Informatica 48~(9)
  (2024).

\bibitem{kanakaraj2024deepn4}
P.~Kanakaraj, T.~Yao, L.~Y. Cai, H.~H. Lee, N.~R. Newlin, M.~E. Kim, C.~Gao,
  K.~R. Pechman, D.~Archer, T.~Hohman, et~al., Deepn4: learning n4itk bias
  field correction for t1-weighted images, Neuroinformatics 22~(2) (2024)
  193--205.

\bibitem{jardim2023image}
S.~Jardim, J.~Ant{\'o}nio, C.~Mora, Image thresholding approaches for medical
  image segmentation-short literature review, Procedia Computer Science 219
  (2023) 1485--1492.

\bibitem{cardenas2018deep}
C.~E. Cardenas, R.~E. McCarroll, L.~E. Court, B.~A. Elgohari, H.~Elhalawani,
  C.~D. Fuller, M.~J. Kamal, M.~A. Meheissen, A.~S. Mohamed, A.~Rao, et~al.,
  Deep learning algorithm for auto-delineation of high-risk oropharyngeal
  clinical target volumes with built-in dice similarity coefficient parameter
  optimization function, International Journal of Radiation Oncology* Biology*
  Physics 101~(2) (2018) 468--478.

\bibitem{west1996comparison}
J.~B. West, J.~M. Fitzpatrick, M.~Y. Wang, B.~M. Dawant, C.~R. Maurer~Jr, R.~M.
  Kessler, R.~J. Maciunas, C.~Barillot, D.~Lemoine, A.~M. Collignon, et~al.,
  Comparison and evaluation of retrospective intermodality image registration
  techniques, in: Medical Imaging 1996: image processing, Vol. 2710, SPIE,
  1996, pp. 332--347.

\bibitem{fitzpatrick2002predicting}
J.~M. Fitzpatrick, J.~B. West, C.~R. Maurer, Predicting error in rigid-body
  point-based registration, IEEE transactions on medical imaging 17~(5) (2002)
  694--702.

\bibitem{marek2018parkinson}
K.~Marek, S.~Chowdhury, A.~Siderowf, S.~Lasch, C.~S. Coffey, C.~Caspell-Garcia,
  T.~Simuni, D.~Jennings, C.~M. Tanner, J.~Q. Trojanowski, et~al., The
  parkinson's progression markers initiative (ppmi)--establishing a pd
  biomarker cohort, Annals of clinical and translational neurology 5~(12)
  (2018) 1460--1477.

\bibitem{marek2024parkinson}
K.~Marek, The parkinson’s progression markers initiative (ppmi)
  clinical-establishing a deeply phenotyped pd cohort am 3.2 (2024).

\bibitem{putcha2023gray}
D.~Putcha, Y.~Katsumi, M.~Brickhouse, R.~Flaherty, D.~H. Salat,
  A.~Touroutoglou, B.~C. Dickerson, Gray to white matter signal ratio as a
  novel biomarker of neurodegeneration in alzheimer’s disease, NeuroImage:
  Clinical 37 (2023) 103303.

\bibitem{carr2002magnetic}
M.~W. Carr, M.~L. Grey, Magnetic resonance imaging: overview, risks, and safety
  measures., AJN The American Journal of Nursing 102~(12) (2002) 26--33.

\bibitem{shellock2000magnetic}
F.~Shellock, Magnetic resonance: safety, bioeffects, and patient monitoring,
  in: Open Field Magnetic Resonance Imaging: Equipment, Diagnosis and
  Interventional Procedures, Springer, 2000, pp. 127--145.

\bibitem{hand2005magnetic}
P.~Hand, J.~M. Wardlaw, A.~M. Rowat, J.~Haisma, R.~Lindley, M.~S. Dennis,
  Magnetic resonance brain imaging in patients with acute stroke: feasibility
  and patient related difficulties, Journal of Neurology, Neurosurgery \&
  Psychiatry 76~(11) (2005) 1525--1527.

\bibitem{xiao2020should}
T.~Xiao, X.~Wang, A.~A. Efros, T.~Darrell, What should not be contrastive in
  contrastive learning, arXiv preprint arXiv:2008.05659 (2020).

\bibitem{robinson2021can}
J.~Robinson, L.~Sun, K.~Yu, K.~Batmanghelich, S.~Jegelka, S.~Sra, Can
  contrastive learning avoid shortcut solutions?, Advances in neural
  information processing systems 34 (2021) 4974--4986.

\bibitem{foo2016progression}
H.~Foo, E.~Mak, T.~T. Yong, M.-C. Wen, R.~J. Chander, W.~L. Au, L.~Tan,
  N.~Kandiah, Progression of small vessel disease correlates with cortical
  thinning in parkinson’s disease, Parkinsonism \& related disorders 31
  (2016) 34--40.

\bibitem{habes2018regional}
M.~Habes, G.~Erus, J.~B. Toledo, N.~Bryan, D.~Janowitz, J.~Doshi,
  H.~V{\"o}lzke, U.~Schminke, W.~Hoffmann, H.~J. Grabe, et~al., Regional
  tract-specific white matter hyperintensities are associated with patterns of
  aging-related brain atrophy via vascular risk factors, but also
  independently, Alzheimer's \& Dementia: Diagnosis, Assessment \& Disease
  Monitoring 10 (2018) 278--284.

\bibitem{dadar2021white}
M.~Dadar, A.~L. Manera, D.~L. Collins, White matter hyperintensities, grey
  matter atrophy, and cognitive decline in neurodegenerative diseases, bioRxiv
  (2021) 2021--04.

\bibitem{debette2010clinical}
S.~Debette, H.~Markus, The clinical importance of white matter hyperintensities
  on brain magnetic resonance imaging: systematic review and meta-analysis, Bmj
  341 (2010).

\bibitem{kim2024tackling}
S.~Kim, C.~Jin, T.~Diethe, M.~Figini, H.~F. Tregidgo, A.~Mullokandov, P.~Teare,
  D.~C. Alexander, Tackling structural hallucination in image translation with
  local diffusion, in: European Conference on Computer Vision, Springer, 2024,
  pp. 87--103.

\end{thebibliography}

\end{document}